\documentclass[preprint,floatfix]{revtex4}
\usepackage[dvips]{graphicx}
\usepackage{epsfig}

\usepackage[cp1251]{inputenc}
\usepackage[english]{babel}
\usepackage{color}
\usepackage{bm}

\usepackage[T1]{fontenc}
\usepackage{xcolor}
\usepackage{lmodern}
\usepackage{listings}
\begin{document}






\title{Non-trivial dynamic regimes of small (nano-scale) quantum systems.}
 
\author{V.A.Benderskii} 
\affiliation {Institute of Problems of Chemical Physics, RAS \\ 142432 Moscow
Region, Chernogolovka, Russia} 

\author{E. I. Kats}  
\affiliation{Landau Institute for Theoretical Physics, RAS, \\
142432, Chernogolovka, Moscow region, Russia.}


\begin{abstract}

Small (but still containing many, about $10^2 - 10^4$, atoms) quantum systems (traditionally termed nano-systems) are dramatically different
from their macroscopic or genuine microscopic (atomic) cousins.  Microscopic molecular systems (with a few atoms) obey a regular quantum dynamics
(described by time dependent Schrodinger equation), whereas in macroscopic systems with continuous energy spectra, one can expect, also regular, although typically relaxation, dynamic behavior. The topic of our paper is in-between these limits. Nano-scale systems are characterized by  small (but finite) mean inter-level spacing. In such a case with recurrence periods in pico-second range, Loschmidt echo  and double resonance phenomena come into the game. System behavior becomes non-trivial and manifests a sort of transitions between regular and chaotic dynamics. 
We show that such dynamic transitions occur when the Loschmidt echo time of life exceeds the typical recurrence cycle period. We illustrate this behavior in the frame work of a few versions of the exactly solvable quantum problem, proposed long ago by Zwanzig. It is based on the study 
of time evolution of the initially prepared vibrational state coupled to a reservoir with
dense spectrum of its vibrational states. In the simplest version of the Zwanzig model, the reservoir has an equidistant spectrum,
and the system - reservoir coupling matrix elements are independent of the reservoir states. We generalize the model to include into consideration the coupling of the initially prepared single state to system phonon excitations. The coupling results to temperature dependent broadening and decay of the echo components. Another generalization is to replace a single level by two states coupled to the Zwanzig reservoir. 
We anticipate that the basic ideas inspiring our work can be applied to a large variety of interesting for the applications nano-systems (e.g., dissipative free propagation of excitations along molecular chains, or as a model for exchange reactions).

\end{abstract}



\maketitle

\section{Introduction}
\label{intro}

There are materials which do not exhibit unusual properties in the nano-scale regime. For example simple non-polar hydrocarbon molecules
when aggregated are added merely additively. In other systems, which we are investigating in this work, the properties may 
exhibit anomalous values (or behavior of a system is unusual) in the nano-regime. Common wisdom borrowed from textbooks on quantum mechanics teaches that a population of somehow initially prepared state of a
macroscopic system, monotonically decreases in time due to energy flow from this initial state into the states of the 
reservoir (formed by all continuous states of the macroscopic system under consideration). In the opposite limit (a small
system with a few degrees of freedom) system behavior is also well known. The system dynamics
is reduced to the recurrence cycles, with their periods (according to the famous Poincare theorem) determined by the lowest rationally
independent inter-level spacings. As it is often the case, an intermediate case (relatively large but not macroscopically large
quantum system) is the most difficult one for theoretical analysis. However just such systems (generically termed as nano-systems) 
containing about several hundreds of atoms, become more and more attractive for various applications.

Our intent in this work is to investigate dynamic behavior of these intermediate nano-size quantum systems. In principle, the full information about system (both dynamics and statics) is naturally contained
in quantum mechanical solution of the corresponding Schrodinger equation. In practice, however the quantum solution is unfeasible even for not
too large (about $10^2 - 10^4$ atoms) systems. Therefore one has to rely either on heavy ab-initio numeric, or to look for analytically doable
approximations. Luckily for us, many years ago Robert Zwanzig in a remarkable work (although published not in a regular and easy accessible
journal \cite{ZW60}) proposed a simple (but not trivial) exactly solvable model of quantum dynamics. Within this model, an initially
prepared single state of a quantum system, which evenly couples to the dense but discrete spectrum of the reservoir levels. Within the Zwanzig model, the coupling strength is assumed to be a constant independent of the reservoir levels. Surprisingly enough that in spite of this evidently erroneous assumption, Zwanzig approximation correctly identifies the dynamic regimes and characteristic time scales in the problem.
Thus, although the Zwanzig model is a toy model (in the sense of caricaturing some physical features), when properly interpreted it can yield quite reasonable values for a variety of essential (and principally  measured) quantities. That such a simple theory can predict rather complex and subtle features for nano-size quantum systems is remarkable.

Here we compile a brief self-contained review that we wish had existed when we first entered the field \cite{BF07}.
In what follows in the frame work of generalized Zwanzig approach we study a number of nano-size quantum systems.
Since our work is primary about physics (and only then how it can be modeled theoretically) it is worth to note first
what are physical systems we have in mind. Our topic is nano-size quantum objects, possessing discrete vibrational energy
spectra. Such objects are carbon nano-tubes, graphene flacks, metallic clusters, some large organic molecules or their clusters.
Typical feature (relevant for everything what follows in the paper) of such systems is that their vibrational spectra
are discrete and characteristic inter-level spacing ${\bar \Delta }$ is on the order of $0.1 \, -\, 10\, cm^{-1}$.
This value of ${\bar \Delta }$ is translated into the pico-second time scale for the periods of the recurrence cycles
\begin{eqnarray}
\label{br1}
T = \frac{2\pi }{c {\bar \Delta }} \simeq 10^{-10} \, -\, 10^{-12}\, s
\, ,
\end{eqnarray} 
where $c$ is the light speed. 
In turn, experimental studies of the excited state evolution in large (nano-size) molecular systems \cite{UM91} - \cite{LE05} 
have established that the processes of vibrational relaxation, energy transfer and ultra-fast chemical reactions occur in the same pico-second time range as recurrence cycles. Therefore, the recurrence cycles are essential ingredients to be included to describe theoretically 
such features of the quantum dynamics of nano-systems as irreversibility, chaotic behavior and loss-free distant energy transfer.   
These and some other, high resolution experiments (which will be cited and shortly discussed in our paper)
demonstrate that a system initial vibrational state evolution, as a rule, has a complicated form with irregular oscillations. 
The methods of double resonances in the non-linear femto-second spectroscopy 
\cite{MU95} - \cite{AS04} 
appear to be especially useful to observe these irregular oscillations of the initial and final state populations simultaneously. The phenomenon has been observed for the wide variety of large-size  molecules in liquid and solid solutions and on interfaces 
\cite{FE03} - \cite{SC02}. Note to the point that irregular oscillations of the population of initially prepared excited state have been confirmed by the numerous quantum dynamical calculations \cite{LW97} - \cite{FZ06}.

The loss-free excitation energy transfer, observed in linear molecules consisting of $12 - 26\, CH_2$ fragments 
\cite{BM98}, \cite{PN01} also belongs to a similar class of irregular in time behavior.  In such kind of experiments the 
initial excitation of the molecular terminal group produces the running wave with approximately constant speed (about of $2\,  ps$ per one fragment). What is even more surprising that instead of naively expected random distribution of the excitation energy over the total set of intra-molecular modes, the wave propagation lasts during several passages along the chain before the uniform distribution is established. The distant energy transfer occurs between localized vibrations of fragments separated by about $1\, nm $ distance during $20\, -\, 40\,  ps$ whereas the life-time of these vibration excitations themselves does not exceed $3\, ps$. The matter is that the excitation energy transfer includes intermediate excitations of delocalized long-lived vibrations binding the initial and final fragments 
\cite{HT03} - \cite{CW09}. 
 
These data permit to suggest that the irregular evolution is the generic property of the systems with dense discrete spectra.
Therefore, the role of the reservoir with a discrete spectrum is fundamentally different from that with a continuous spectrum, typical for
macroscopic systems. In the latter case (with the infinite recurrence period time) the reservoir serves as a sink for energy flow. Contrary  in the case of the discrete spectrum, repetitive reverse transitions from the reservoir to the initial state and in the opposite direction, determine the non-trivial and often irregular long-time dynamics. The synchronization of these reverse transitions results in the appearance of a multi-component Loschmidt echo phenomenon with a partial recovery of the initial state population (at the frequency corresponding to the initial excited state energy), and double resonances (at the frequencies of the reservoir states). Both effects are responsible for non-monotonic time evolution. The counterpart of such behavior is the periodic energy concentration in one of the vibration mode, arising as a result of the time dependent exchange between reservoir states far from equilibrium. 

In a few of previous works of our group 
\cite{BF07}, \cite{BG09} - \cite{BK11b} we illustrated how such complex behavior might appear in the frame work of the simple Zwanzig
model, and how the simplest version of the model can be generalized to relax some unphysical assumptions of the model. Besides in these works we proposed a method which makes it possible to solve the dynamical problem analytically beyond the bare Zwanzig model approximations (
\cite{BG09b} - \cite{BK11b}). 
The main ingredients of our new method (only schematically and briefly described in the previous papers) is the representation of the partial amplitudes of recurrence cycles. Unlike the standard Fourier expansion over eigen-frequencies, this representation reveals explicitly the time dependent exchange between intra-molecular states in each recurrence cycle. The fine structure of the Loschmidt echo arises as a result of the dephasing phenomena associated with the fact that the exchange of the different reservoir states with the initial state occurs not at the same instants of time. The synchronization of the reverse transitions is destroyed when the Loschmidt echo components of the different recurrence cycles start to overlap. The cycle overlapping determines the critical recurrence cycle number. Then at larger time (cycle numbers) the system dynamics is expected to be in the stochastic-like regime.

Since that time we have realized that the quantum dynamics of nano-size systems  is much richer than that predicted for the systems with continuous spectra, and that non-monotonous in time and irregular dynamics is a robust and generic
feature of almost arbitrary quantum system with $10^2\, -\, 10^4$ degrees of freedom. Moreover the regular-stochastic dynamic transition 
(crossover) yields to a loss of the expected one-to-one correspondence between system spectrum and its long time dynamics. For all our examples the spectrum remains deterministic, while long time dynamics eventually becomes stochastic. This unusual combination is the specific feature of nano-systems. 

Motivated by this new understanding we decided to combine altogether our previous works to illustrate our method on a number of 
particular physical realizations (only partially overlapping with those in the previous works). The aim to present this review arises from the conviction that unifying our previous works supplemented
by the new applications of the developed theoretical approach and 
by new experimental data and observations collected in very recent year, yield to a new stage of development of this field: dynamics of nano-systems.

Our manuscript is divided  into 7 sections. After this introduction in section \ref{zwanzig}  we summarize shortly  the results on quantum dynamics of the bare Zwanzig model. It is discussed also a possible physical realization most closely satisfying the model assumptions. The section contains also an extended list of references to compensate partially its brevity.
Then in section \ref{initial} utilizing developed in our works \cite{BG09b} - \cite{BK11b} theoretical approach we investigate with all details the evolution of the population of the initially prepared single state of the system. 
In section \ref{generalizations} we analyze various physically motivated generalizations of the bare Zwanzig model, and how our analytical
method should be modified to describe theoretically these generalizations. In the same section (subsection \ref{ensemble} we show that on similar footing we can study dynamics not only for an individual (single) nano-system but as well for the ensemble of somehow distributed nano-systems.  We investigate the reservoir states evolution in section \ref{reservoir}. We collect some already discussed in the literature and new applications of our approach to physically interesting phenomena in nano-systems. Namely, two-level systems coupled to reservoir (subsection \ref{two}),
propagation of vibrational excitations in nano-size chains (subsection \ref{propagating}). 
The last section
\ref{con} summarizes the main findings of our work, with a discussion of possible physical consequences and interpretation of the results.

\section{Zwanzig model and its approximate physical realization}
\label{zwanzig}

Mentioned in the previous section Zwanzig model \cite{ZW60} considers a quantum system where somehow initially
prepared a single excited vibrational energy level is coupled with a constant coupling strength ($C$) to the equidistant discrete reservoir
states. In such a formulation (termed in what follows as the bare Zwanzig model) the model allows the exact
analytic solution. We will present below briefly the exact solution (for time evolution of the population for the initial state and reservoir states) found in \cite{BF07}, however it is worth to noting first about a possibility of physical realizations of the bare Zwanzig model
assumptions. As an approximate realization one can have in mind the well-known scheme of radiation-less transitions \cite{UM91}.
In this experimentally realizable scheme the initially excited state is prepared by a short optical pulse pumping (see Fig.1). 

\begin{figure}
\begin{center}
\includegraphics[width = 225bp]{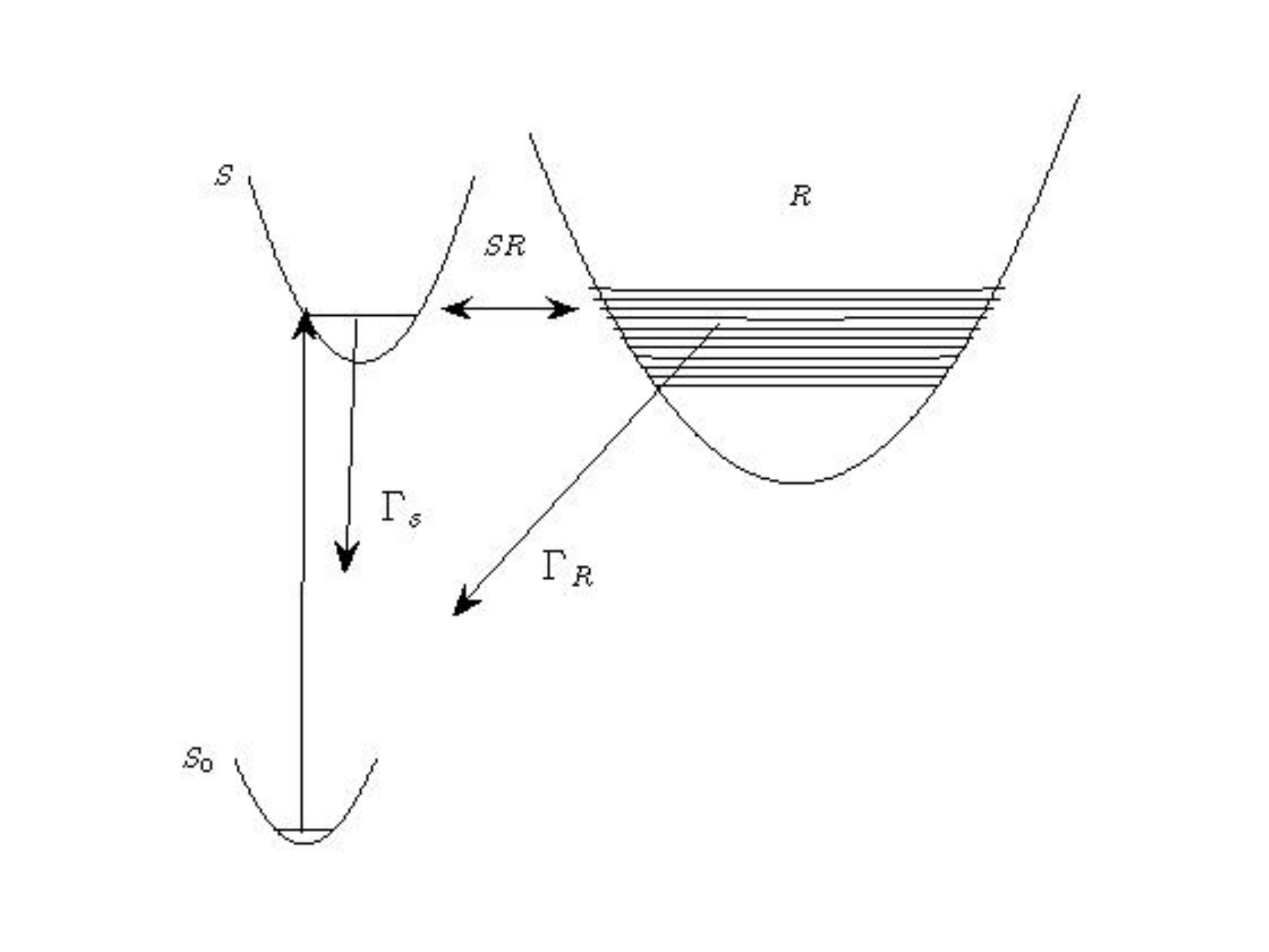}
\caption{Scheme of  radiation-less transitions $\{R_n\} \leftarrow S$, accompanying by the optical transition from the ground to initial state $S \leftarrow S_0$. Direct optical transitions $\{R_n\} \leftarrow S_0$  are forbidden.}
\label{1}
\end{center}
\end{figure}

It is essential that the initial vibrational state $S$ belongs to the so-called narrow potential energy surface. If the intervals between neighboring levels on this surface exceed the spectral width of the pumping pulse, only a single initial state enters into the game. The final states $\{R_n\}$ belong to the bright low-lying energy surfaces, but the direct optical transitions to these states are forbidden (e.g., due to symmetry restrictions). The $S R$ interaction leads to a formation of the mixed (coupled) states, which are certain superpositions of the initial and reservoir states. The absorption band related to the transition from the ground to these mixed states acquires the fine structure with 
the components which are determined by the individual mixed states. The envelope of the absorption band $S \leftarrow S_0$  is determined by the width of the $S R$ interaction region. Since the reservoir states are the eigen-states of the single energy surface, they are ortho-normal states. The Hamiltonian describing the $S R$ interaction between the initial state (with its eigen-function $\varphi _s$) and the reservoir states (with their eigen-functions $\{\varphi _n\}$   has the following form (in the units where $\hbar = 1$)

\begin{eqnarray}
\label{br2}
H = \varepsilon ^0_s b_s^+ b_s + \sum _{n} \varepsilon _n^0 b_n^+ b_n + \sum _{n}
C_n( b_s^+ b_n + b_s b_n^+)
\, ,
\end{eqnarray} 
where $b_s^+ , b_n^+ , b_s , b_n $  are creation and annihilation operators of the corresponding states
\begin{eqnarray}
\label{br3}
b_{s, (n)}\Phi _0 =0\, ,\, b_{s, (n)}^+ \Phi _0 = \varphi _{s, (n)}\, ,\, b_{s (n)}\varphi _{s (n)} = \Phi _0\, ,\,
b_{s (n)}^+ \varphi _{s (n)} =0
\, ,
\end{eqnarray} 
with $\Phi _0$ being the ground state wave function. The last of the Eqs (\ref{br3}) is valid if the resonance region of the $S R$ interaction contains only one excited state per each vibrational mode. 
For the bare Zwanzig model Hamiltonian the coupling constant $C_n$ is independent of $n$, i.e., $C_n \equiv C$.
Since the basis set of the reservoir states is ortho-normal, the Hamiltonian matrix has only one non-zero column and one non-zero line of the interaction matrix elements (apart from the diagonal elements corresponding to the bare (non-perturbed by the interaction) eigen-values. Owing to this simple structure of the Hamiltonian matrix, the secular equation reads as  
\begin{eqnarray}
\label{br4}
F(\varepsilon ) = \epsilon \sum _{n}\frac{C_n^2}{\varepsilon - \varepsilon _n^0} = 0
\, ,
\end{eqnarray} 
where the energy of the initial state is chosen as the zero, and the inter-level spacing is chosen as the unity of the energy scale.
The dipole matrix element of the $S \leftarrow S_0$ transition is determined by the energy presentation of the Green function  (see e.g., 
\cite{UM91} or its textbook version in \cite{MA76}. 
Within the above assumptions (the single initial state and forbidden direct $\{R_n\} \leftarrow S_0$
transitions), the dipole moment $\mu $ of the transition is energy independent, and the absorption band is determined
by the following function $I(\varepsilon )$
\begin{eqnarray}
\label{br5}
I(\varepsilon ) = - \pi ^{-1}|\mu |^2 Im \langle \varphi _s |G(\varepsilon )|\varphi _s \rangle
\, ,
\end{eqnarray} 
where
\begin{eqnarray}
\label{br6}
G(\varepsilon ) = (\varepsilon - H)^{-1}
\, .
\end{eqnarray} 
Since the poles of the Green function $G(\varepsilon )$ 
in the lowest half-plane are defined by the roots of the secular equation 
\begin{eqnarray}
\label{br66}
lim _{\eta \to 0}\langle \varphi _s|G(\varepsilon + i\eta)|\varphi _s\rangle = (F(\varepsilon ))^{-1}
\, ,
\end{eqnarray} 
and the function $1/F(\varepsilon )$ in the vicinity of the pole  $\varepsilon = \varepsilon _n^*$ is proportional  to the residue in this pole, the shape of the absorption band  is given by
\begin{eqnarray}
\label{br7}
\langle \varphi _s |G(\varepsilon )|\varphi _s\rangle \simeq - i \pi \sum _n \left . \left (\frac{dF}{d\varepsilon }\right )^{-1} \right |_{F(\varepsilon )} = 0
\, .
\end{eqnarray} 
The sum in Eq. (\ref{br4}) is singular in all points $\varepsilon = \varepsilon _n^0$,
so that one root is located in each interval  $[\varepsilon _n^0\, ,\, \varepsilon _{n+1}^0$. 
Separating the singular term $(u_n^{-1})$ from the bare energy value $\varepsilon _n^0$, (i.e., $\varepsilon _n = \varepsilon _n^0 + u_n$) one can rewrite Eq. (\ref{br4})  as the expansion over the moments of the mixed reservoir states
\begin{eqnarray}
\label{br8}
F(u_n) = u_n + \varepsilon _n^0 - \frac{C_n^2}{u_n} + \sum _{\nu =0}^{\infty } M_{n \nu }u^\nu 
\, .
\end{eqnarray} 
Here 
\begin{eqnarray}
\label{br9}
M_{n \nu } = \sum _{k}\frac{C_{n+k}^2}{\Delta _{n+k}^{\nu + 1}}\, ;\, \Delta _{n +k} = \varepsilon _{n+k}^0 - \varepsilon _n^0
\, .
\end{eqnarray} 
In the simplest bare Zwanzig model \cite{ZW60}, the spectrum of the unperturbed reservoir is assumed to be equidistant and interaction matrix elements $C_n$ and line widths $\gamma _n$ are the same for the all reservoir states, i.e., 
\begin{eqnarray}
\label{br10}
\varepsilon _n^0 = n\, ,\, C_n = C\, ,\, \gamma _n = \gamma
\, .
\end{eqnarray} 
Then, the Eq. (\ref{br4}) reads as 
\begin{eqnarray}
\label{br11}
F(\varepsilon ) = \varepsilon + i \gamma - \Gamma cot (\pi (\varepsilon + i \gamma )) =0
\, ,
\end{eqnarray} 
where for the sake of simplicity, the reservoir level $n=0$  is assumed to be in the resonance with the initial state. 
Eqs. (\ref{br5}), (\ref{br7}), and (\ref{br11}) show that the envelop of the absorption band is the Lorentzian and its half-width is equal to the interaction region width $\Gamma $
\begin{eqnarray}
\label{br12}
\rho (\varepsilon ) = \pi ^{-2}\frac{\Gamma }{\varepsilon ^2 + \Gamma ^2} \sum _{n}\frac{\gamma }{(\varepsilon - n)^2 + \gamma ^2}
\, .
\end{eqnarray} 
where $\Gamma = \pi C^2$.
To get the well-resolved spectrum, the half-widths of the components $\gamma _n$ has to be smaller than the characteristic inter-level
spacing 
\begin{eqnarray}
\label{br13}
\gamma _n \ll \varepsilon _{n+1} - \varepsilon _n
\, .
\end{eqnarray}

\section{Initial state evolution within the bare Zwanzig model.}
\label{initial}
Textbook quantum mechanical wisdom teaches that an arbitrary wave function is expanded over the complete basis set of the eigen-functions with time dependent coefficients
\begin{eqnarray}
\label{br14}
\Psi (t) = a_s(t)\varphi _s(q) + \sum _{n} a_n(t)\varphi _n(q_n)
\, ,
\end{eqnarray} 
where $q$ and $q_n$ stand for the initial state and reservoir states degrees of freedom.
These coefficients satisfy the Heisenberg equations of motion
\begin{eqnarray}
\label{br15}
i\frac{\partial a_s}{\partial t} = \epsilon _s^0 a_s + \sum _{n} C_n a_n\, ;\,
i\frac{\partial a_n}{\partial t} = \varepsilon _{n}^0 + C_n a_s
\, .
\end{eqnarray} 
Supplementing the Eqs. (\ref{br15}) by the natural initial conditions
\begin{eqnarray}
\label{br16}
a_s(0) = a_{s0}\, ,\, a_n(0) = a_{n0}
\, 
\end{eqnarray} 
their solution can be expanded into Fourier series. The terms in this series are determined by the energy eigen-values of the Hamiltonian matrix, i.e., by the secular equation (\ref{br4})
\begin{eqnarray}
\label{br17}
a_s(t) = (2\pi i)^{-1} \int _{-\infty }^{\infty }(a_{s0} + R(\varepsilon ))\exp (-i \varepsilon t)\frac{d \varepsilon }{d F(\varepsilon )} =
\sum _{n}(a_{s0} + R(\varepsilon ))\left .\left (\frac{d F}{d\varepsilon } \right )\right |_{F(\varepsilon ) = 0}
\, ,
\end{eqnarray} 
where the entering (\ref{br17}) function $R(\varepsilon )$ is
\begin{eqnarray}
\label{br18}
R(\varepsilon ) = \sum _{n}\frac{C_n a_{n0}}{\varepsilon - \varepsilon _n^0}
\, .
\end{eqnarray} 
For the bare Zwanzig model and zero value for the initial amplitudes of all reservoir states, $a_{n0} =0$, the time-dependent initial state amplitude is given by 
\begin{eqnarray}
\label{br19}
a_s(t) = 2 \sum _{n=0}^{n=\infty }\frac{\cos (\varepsilon _n t)}{1 + \pi ^2 C^2 (\varepsilon _n/C)^2}
\, .
\end{eqnarray} 
The sum (\ref{br19}) gives the exact solution of the dynamical problem. If the initial state is strongly coupled to reservoir states
($\Gamma \gg 1$), the sum contains the large number of terms oscillating with rationally independent frequencies. It is not an easy task
to identify the recurrence cycles from such a sum of the oscillating terms. To overcome the difficulty we apply the generalized Poisson summation formula introduced in \cite{BG09}, \cite{BG09b}. 
This method makes it possible to express the initial state amplitude $a_s(t)$ in terms of the recurrence cycle partial amplitudes $a_s^{(k)}$
\begin{eqnarray}
\label{br20}
a_s(t) = \sum _{k}a_s^{(k)}(t - 2k \pi )
\, 
\end{eqnarray} 
with the partial cycle amplitudes expressed as
\begin{eqnarray}
\label{br21}
a_s^{(k)}(t - 2 k\pi ) = \frac{\Gamma }{\pi } \exp (-2k \pi \gamma )\int _{-\infty }^{\infty } d\varepsilon 
\frac{(\varepsilon + i (\Gamma - \gamma _s))^{k-1}}{(\varepsilon - i(\Gamma + \gamma _s))^{k+1}}
\exp (i\varepsilon (t - 2 k\pi ))
\, ,
\end{eqnarray} 
where $\gamma _s$ and $\gamma $ are the widths of the initial and reservoir states respectively. 
Pure mathematically Eq. (\ref{br21}) replaces a discrete set of poles situated close to the real axis by a single pole on the imaginary axis, which characterizes the decay probability of the initial quasi-stationary state.  

We are interested in the initial state population evolution for $t \geq 0$. Then the sum (\ref{br20}) contains only the finite number of partial amplitudes for the cycle numbers in the interval $0 \leq k \leq [t/2\pi ] $ ($[.]$ denotes the integer part). 
For the $k=0$ cycle, the integrand the in Eq (\ref{br21}) has two poles at $\varepsilon = \pm i \Gamma $. Only the positive pole is inside the integration contour comprising the real axis and half-circle of an infinite (very large) radius in the upper complex half-plane, where the integrand is zero. For $k <0$ only the negative pole remains. and it is outside the integration contour. Hence, the cycles with $k<0$  do not contribute to
$a_s(t)$. For $k > [t/2\pi ]$, the integration contour lies in the lower complex  half-plane and no pole $\varepsilon = i \Gamma $. Therefore $a_s^{(k)}$ are nonzero only for $t \geq 2 k \pi $, i.e., they proportional to the Heaviside step functions $\theta(t - 2 k\pi )$.
In the initial cycle $k=0$, the amplitude, describing the initial state $S$ population evolution, exhibits exponential decay, with
its decrement including that of the intrinsic (own) energy level $S$ decay and the decrement due to $S - R$ -transitions
\begin{eqnarray}
\label{br22}
a_s^{(0)}(t) = \exp (- (\Gamma + \gamma _s)t)
\, .
\end{eqnarray} 
The strong coupling condition $\Gamma \gg 1$  ensures exponential decay when the $S R$ -transitions occur into the large number of the final states of the reservoir with $n \leq \Gamma $. Then the wave function destructive  interference suppresses coherent oscillations.  
In the weak coupling limit, resonance transitions between the initial state $S$ and the reservoir state $n=0$ govern the dynamic behavior. 
It is worth to note to the point that the decay rate predicted by the Fermi Golden Rule approximation is $2\Gamma $. The reason for the error is obvious: this approximation is valid only if the amplitude is a slow (in comparison with the rapidly oscillating terms) function 
of time, what is evidently not the case for the essential oscillating contributions of the recurrence cycles in the strong coupling case 
\cite{BK08}. All the more the subsequent revivals of the initial state amplitude for the cycles $k \geq 1$ can not be reproduced within the Fermi Golden Rule approximation.
 
Calculating residue in the $(k+1)$-th order pole at $\varepsilon = i(\Gamma + i \gamma _s)$, one can express the integral (\ref{br21}) in terms of the ortho-normalized set of the Laguerre functions (the products of the Laguerre polynomials and exponential factors) 
\cite{BG09}, \cite{BG09b}
\begin{eqnarray}
\label{br23}
a_s^{(k)}(t) = \exp (- 2k\pi \Gamma t) \exp (-(\gamma _s (t - 2k\pi )))
\, ,
\end{eqnarray} 
where $a_{s0}^{(k)}(t)$ is the partial amplitude at $\gamma = \gamma _s =0$ 
\begin{eqnarray}
\label{br24}
a_{s0}^{(k)}(t) = 
\frac{\tau _k}{k}\frac{dL_k(\tau _k)}{d\tau _k}\exp (- (\tau _k/2))\theta (\tau _k) 
= - \frac{\tau _k}{k}L^1_{k-1}(\tau _k) \exp (-(\tau _k/2))\theta (\tau _k)
\, ,
\end{eqnarray} 
where $\tau _k$ stands for the local time of the $k$ -th cycle
\begin{eqnarray}
\label{br25}
\tau _k = 2 \pi (t - 2k\pi )C^2
\, .
\end{eqnarray} 
The expressions (\ref{br20}) - (\ref{br25}) above present the exact solution of the dynamical problem for the bare  Zwanzig model. 
Note that the initial state evolution can be described by two equivalent sums (\ref{br19}) and (\ref{br20}). The first sum being  the Fourier transform of the energy spectrum, presents the initial state amplitude as a superposition of the coherent oscillations, whereas the second sum depending on the total transition probability into all reservoir states,  manifests explicitly the partial recovery of the initial state amplitude during each recurrence cycle. This phenomenon is known as Loschmidt echo. From the expressions (\ref{br23}) - (\ref{br25}) we
see the following properties of the Loschmidt echo: (i) the number of zeros and the number of components of $k$-th partial amplitude is
$k$, (ii) the total width of the state population evolution in time (which is determined by the width of the oscillation region for the Laguerre polynomials) is $4k$. We illustrate these properties in Fig. 2, where we show the results of the direct Matlab solutions 
of the Eqs. (\ref{br20}) - (\ref{br25}).
\begin{figure}
  \begin{center}
    \includegraphics[height=3.5in, width = 225bp]{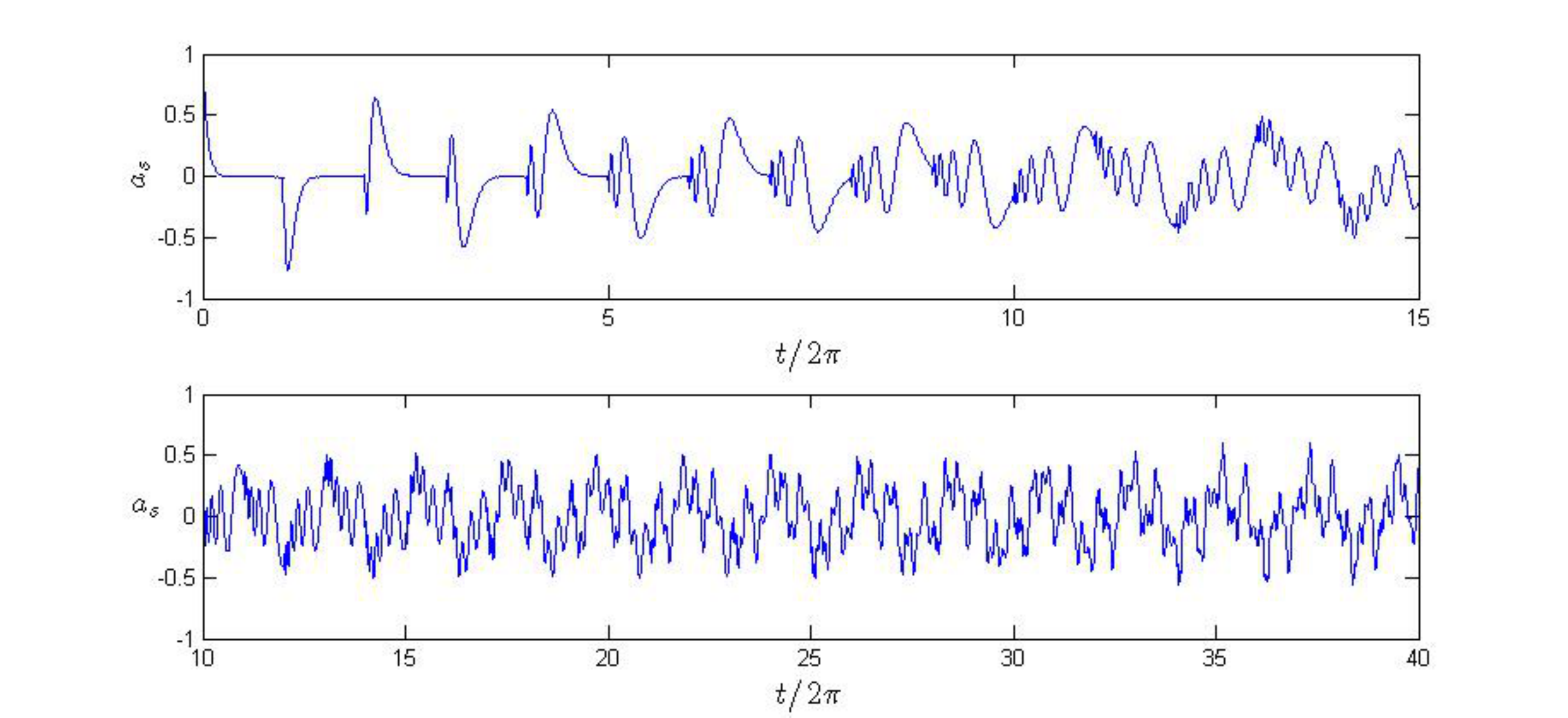}
    \caption{The initial state evolution for the bare Zwanzig model. $C^2 = 1$, and $\gamma = 0$ The upper panel shows the initial cycles with the regular dynamics. The lower panel demonstrates chaotic-like long time dynamics in the case of the overlapping recurrence cycles.}
    \label{2}
  \end{center}
\end{figure}
One essential physical remark is in order here. The fact is that since   
the recurrence cycle period increases proportionally to density of the reservoir states. Eventually it becomes longer than the time interval available for measurements. It happens when the density of states exceeds the corresponding particular experiment-dependent limit. In this case only the exponential decay in the cycle $k=0$ is observed. This limit, (depending on the measurement resolution), distinguishes the systems with continuous and discrete spectra. As we emphasize in the Introduction section, nano-particles posses typically the discrete vibrational spectra. For such a system during the cycle $k=0$ the initial state population is completely transferred to the reservoir states. 
At the beginning of the cycle $k=1$, the reverse transitions partially restore the initial state amplitude, $a_s(t)$. However at the end of the first cycle the initial state population becomes to be depleted again. Since the $S R$  exchange of the populations leads to the additional phase shifts, the amplitudes of the reservoir states at the beginning of second cycle $k= 2$ consists of two components with different phases, and their exchange with the initial state produces the two-component echo signal. The first its component is related to amplitude  that comes from the cycle $k=1$, while the second component appears due to the $S R$ exchange in the cycle $k=2$. Similarly, the partial amplitude of the $k$ -th cycle split into  $k$ components with different phases coming from the preceding cycles. Since the total width of the echo signals grows with increasing of the cycle number, the components of the neighboring cycles begin to overlap when the cycle width becomes equal to the cycle period, $4\pi \Gamma $. This critical cycle number is
\begin{eqnarray}
\label{br26}
k_c = \pi ^2 C^2
\, .
\end{eqnarray} 
For $k < k_c$ the cycles do not overlap and the Loschmidt echo signal has the well-resolved components. For $k > k_c$  the echo signal structure becomes more and more complex due to mixing of the components of the different cycles. 
The specific features of the above described evolution, can be formulated and analyzed in terms of the Langevin-like dynamic  equations.
Indeed, eliminating the amplitudes of the reservoir states, the Heisenberg equations of motion 
(\ref{br15}) can be written as those for the harmonic oscillator with dissipative (friction) and reactive forces
\begin{eqnarray}
\label{br27}
{\ddot{a}_{s0}} + C^2 \int _{0}^{t} {\dot{a}_{s0}}(t^\prime ) G(t - t^\prime ) dt^\prime = - C^2 G(t) \, ;\,
G(t) = 2\pi \sum _{k=-\infty }^{\infty }\delta (t - 2k\pi ) - 1
\, .
\end{eqnarray} 
The Eqs. (\ref{br27}) show that the transitions to the resonance level $n=0$ lead to the coherent oscillations of the
initial state population with the frequency which is determined by the $S R$ interaction matrix element.
The transitions to the non-resonant levels $n \neq 0$ produces the friction and reactive forces. In the bare Zwanzig model, the reactive force has the form of the periodic $\delta $ -pulses with the recurrence cycle periodicity. 
It is instructive to note the difference in the dynamic behavior between the Zwanzig model and the
extensively studied models for the kicked periodically systems (see e.g., the monographs \cite{ZA85}, \cite{TA89}, and reviews works \cite{FH96}, \cite{IZ90}, and
the literature cited therein). The matter is that Within the Zwanzig model the friction force (the integral term in the Eq. (\ref{br27}) is non-local and it depends on the initial state amplitude. Besides the reactive force has a fine structure and it contains the infinite number of the periodic terms.

The average value of the initial state population per cycle can be expressed as
\begin{eqnarray}
\label{br28}
\langle \left (a_{s0}^{(k)}\right )^2 \rangle _k = \frac{\pi }{2 k k_c}\int _{0}^{4k_c}|a_{s0}^{(k)}(\tau _k)|^2 d\tau _k
\, . 
\end{eqnarray} 
The integral (\ref{br28}) is $k$-independent and in the both limits $k < k_c$ and $k > k_c$ it is approximately $1/\Gamma $. 
The constant average value for cycle amplitudes in the limit $k \gg k_c$ signals that the mixing in the strongly overlapped cycles is a stationary random process. Owing to the cycle overlap the dynamics includes the large number of combination frequencies. Their interference results in the stochastic behavior of the time dependent amplitudes. 
The stationary behavior is confirmed also by the constant number of the extremal points  per cycle ($\simeq 2 k_c$) and is $k$ -independent decay of the average per cycle correlation functions. These properties are typical for a dynamical system with mixing.
Such kind of the paradoxical behavior (irregular, quasi-chaotic dynamics of the
deterministic systems) admits a statistical description. The characteristics of the bare Zwanzig model, found above, show that this model can also be treated as a deterministic dynamical system whose long-time evolution is a random-like. Mixing of zeros of the state population amplitudes, and decrease in the intervals between these zeroes make the initial state amplitude very sensitive to small variations of the time.
Therefore an arbitrary small coarse graining destroys the deterministic dynamics. Upon such coarsening, the system loses both the reversibility in time (i.e., the initial state wave function cannot be determined from the long-time evolution of the coarsening amplitudes) and one-to-one correspondence between the system dynamics and its spectrum. In order to retain the deterministic character of the long-time evolution, it is necessary to measure more and more rapidly oscillating amplitudes within more and more narrow time intervals, which is impossible in practice for the finite measurement accuracy. One can say that the chaos appears due to a coupling of the dynamical system to a measurement device
(c.p., with \cite{ZU82} or \cite{GR93}).

\section{Generalizations of the Zwanzig model}
\label{generalizations}

The first step to relax the rather severe restrictions of the bare Zwanzig model, is to deform somehow the reservoir equidistant 
spectrum. To illustrate how it works in the following subsection \ref{simple} we study only two special types of the spectrum deformations,
and then, armed by this knowledge in subsection \ref{arbitrary} we investigate more general spectral deformations, enabling to describe
almost arbitrary reservoir spectra. It is worth to note that if the order of the reservoir levels does not change upon deformations
$(\varepsilon _{n+1}^0 - \varepsilon _n^0 > 0)$ (i.e., for any $n$, there are no level permutations), the spectrum remains regular.
If it is not the case, mixing of the levels occurs. One should distinguish this spectral mixing, producing chaotic behavior in quantum mechanic
systems, and also known dynamic mixing phenomena leading to chaotic classical dynamics \cite{SI77} - \cite{HA90}.

\subsection{Simple deformations of the equidistant reservoir spectrum}
\label{simple}

To illustrate how it works in this subsection we study only two special types of the spectrum deformations. Namely,
\begin{itemize}
\item
Homogeneous spectral deformations, producing from the bare Zwanzig model reservoir, that with non-equidistant but still regular spectrum
\begin{eqnarray}
\label{br29}
\varepsilon _n^2 = n^2(1 + a^2 n^2)^{\pm 1};\, C_n^2 = C^2(1 + b^2 n^2)^{\pm 1}
\, , 
\end{eqnarray} 
where the sign $\pm $ correspond to stretching (sign $+$) or compression (sign $-$) of the spectrum, and constants $a$ and $b$
are assumed to be small $a\, ,\, b\, \ll 1$.
\item
Splitting of the spectrum into $K$ displaces sub-lattices with the same periods for each of the sub-lattice
\begin{eqnarray}
\label{br30}
\varepsilon _{n0} = \pm n K\, ;\, \varepsilon _{nk} = \pm (n K + k) + x_k
\, , 
\end{eqnarray}
where $k = 0\, ,\, 1\, ,\, K-1$ and $x_1 < x_2 ...<x_{K-1} < 1/2$.  
\end{itemize}
Spectral mixing occurs when the reservoir spectrum is split into sub-lattices with rationally independent periods,
i.e.,
\begin{eqnarray}
\label{br31}
\varepsilon _{n0} = \pm n K\, ;\, \varepsilon _{nk} = \pm (n K + k)(1 + \delta _k)
\, , 
\end{eqnarray}
where $k = 0\, ,\, 1\, ,\, K-1$ and $\delta _1 < \delta _2 ...<\delta _{K-1} < 1/2$.  
The deformation (\ref{br31}) gives to the resonances between the levels from the different sub-lattices at the values of $n k$ and
$n^\prime k^\prime $  leading to small denominators in the expansion (\ref{br8}). As a result, the tails of the envelope
of the absorption band are deformed in the far spectral region for the reservoir levels with $n \simeq 1/\delta _k$ (although the absorption band central region is almost unchanged under this spectral deformation).

In order to study the absorption band shape for an arbitrary reservoir spectrum, it is convenient to utilize the formal scaling relation
transforming the arbitrary spectrum into the equidistant one. One more advantage of this scaling transformation that it allows us to discuss 
the spectral and dynamical effects of deformations on the same footing. For the both types of the non-equidistant spectra listed above
the secular equation has the form
\begin{eqnarray}
\label{br32}
F(\lambda ) = P(\lambda )(Q(\lambda ) - cot (\pi \lambda )) = 0
\, , 
\end{eqnarray}
where the auxiliary variable $\lambda $  is introduced in order to map the sub-sequence of the intervals $[\varepsilon _n^0\, \varepsilon _{n+1}^0]$ for the reservoir spectrum under consideration into the sub-sequences $[n\, n+1]$ for the bare Zwanzig model
reservoir levels. To perform the mapping we introduce two functions $g(\lambda )$ and $h(\lambda )$ which determine the deformation of the spectrum
\begin{eqnarray}
\label{br33}
\varepsilon _n^0 = ng(\lambda ) \delta (\lambda - n)\, ;\, C_n^2 = C^2 h(\lambda ) \delta (\lambda  - n)
\, .
\end{eqnarray}
The summation in the secular equation (\ref{br4}) for the deformed spectrum gives 
\begin{eqnarray}
\label{br34}
P(\lambda ) = \Gamma \frac{h(\lambda )}{g(\lambda )}
\, .
\end{eqnarray}
The existence of small or even vanishing inter-level spacings in the generalized secular equation (\ref{br32}) is related to the characteristic function $Q(\lambda )$,  which in own turn  depends on $\Gamma $ and the deformation parameters entering the Eqs. 
(\ref{br29}) - (\ref{br31}).
For mixing-free deformations the moments $M_{\nu n}$ (\ref{br9}) are monotonic functions of the deformation parameters. However when
the deformation parameter $\delta $ entering in the  Eq. (\ref{br31}) grows upon certain critical value. at least one interval
approaches to zero. 
These very small intervals correspond to the maxima of the $M_{\nu n}$ moments.
The number of small intervals increases upon increasing of the mixing  parameter, $\delta $.  Moreover unlike the homogeneous deformations
(\ref{br29}) and (\ref{br30}) which decrease the number of mixed states, the mixing deformation (\ref{br31}) leads to a broadening
of the absorption band. The critical value of the deformation parameter $\delta $ corresponds to the appearance of one permutation of order of the intervals within the interaction region, $\Gamma $. If the number of mixing sub-lattices is three ($K=3$), the critical value is
$\delta _c \simeq (1/3\Gamma )$. Our numerical computations suggest the following relation between the critical deformation parameter and
$\Gamma $, namely 
\begin{eqnarray}
\label{br35}
\delta _c \Gamma \simeq 0.20\, -\, 0.35
\, .
\end{eqnarray}
Eq. (\ref{br35}) is valid for two-parametric deformations including superposition of homogeneous deformations with mixing. 
When the deformation parameters grow, the isolated resonances start to overlap  and initially regular spectra look as random-like ones. 
The results of our numeric computations, presented in Fig. 3 illustrate this phenomenon (crossover from the regular to quasi-random spectrum upon increasing the deformation parameter).
\begin{figure}
  \begin{center}
    \includegraphics[height=3.5in, width = 225bp]{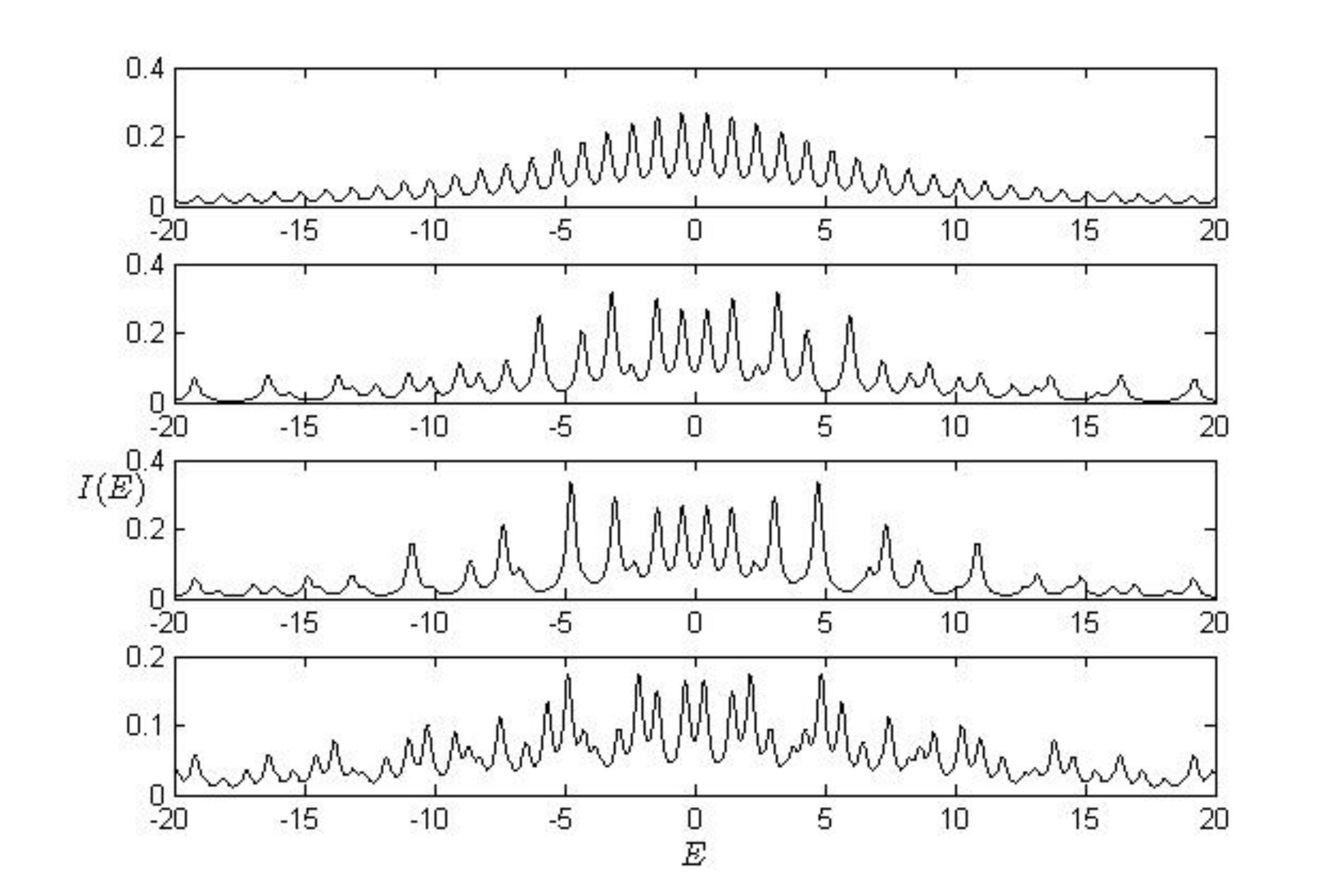}
    \caption{Crossover from the regular to quasi-random spectrum of the mixed $R S$ states upon the mixing deformation.
$K=3$, $\delta _1 = 0\, ;\, 0.1\, ;\, 0.4\, ,\, 0.7$, $\delta _2 = 2\delta _1$ (from the top to the bottom). $C^2 = 2\gamma = 0.1$.}
    \label{3}
  \end{center}
\end{figure}
It is worth to note that this figure shows that the duration of the transition from the beginning of overlapping of the recurrence cycles 
to deep mixing behavior is several times longer than the overlapping time. The wide crossover region, (''weak chaos'') is due to the multicomponent structure of the Loschmidt echo, where oscillations with different frequencies and phases are involved. The influence of the reservoir spectrum on the duration of the transition region is manifested in the ''lag effect'' discussed in Section \ref{reservoir}.

\subsection{Arbitrary reservoir spectra.}
\label{arbitrary}
To analyze more general deformations of the reservoir spectra, let us re-write the expression for the partial recurrence cycle
amplitudes (\ref{br24}) in the form valid for the deformed reservoir spectrum. For the secular equation (\ref{br32}), 
the partial amplitudes can be derived using as above the generalized  Poisson summation (which works even for the complex-valued
$\lambda $) we get
\begin{eqnarray}
\label{br36}
a_s^{(k)}(\tau ) = \pi ^{-1} \int _{-\infty }^{\infty }\exp (i(P(\lambda )Q(\lambda )t - 2k\pi \lambda ))\frac{d\varepsilon }{d \lambda }
\frac{1}{P(\lambda )(1 + Q^2(\lambda ))}\left (\frac{Q(\lambda ) + i}{Q(\lambda ) - i}\right )^k d \lambda 
\, .
\end{eqnarray}
For the bare Zwanzig model, $P(\lambda )=1$, $Q(\lambda ) = \lambda $, and $\lambda = \varepsilon /\Gamma $. The main new feature (in comparison to its Zwanzig model counterpart (\ref{br21})) of the expressions (\ref{br36}) for the partial cycle amplitudes,  
derived from the generalized secular equation (\ref{br32}),  is that the recurrence cycle period in Eq. (\ref{br36}) depends on time and, hence, local time of  $k$-th cycle depends on $\varepsilon /\lambda $
\begin{eqnarray}
\label{br37}
\tau _k(\lambda  ) = \frac{\varepsilon }{\lambda }t - 2k\pi 
\, .
\end{eqnarray}
The deformations change the number and positions of the poles in the integrand in Eq. (\ref{br36}). Therefore the integration contour
should be also changed to be adapted to the spectrum deformations. 
If the deformation has a form
\begin{eqnarray}
\label{br38}
Q(\lambda ) = \lambda + \eta \phi  
\, ,
\end{eqnarray}
where $\eta \ll 1$, and $\phi (\lambda )$ is assumed to be a single power function
\begin{eqnarray}
\label{br39}
\phi (\lambda ) = \lambda ^{2s + 1}\, ,\, s \geq 1 
\, .
\end{eqnarray}
For such a deformation in the integrand in Eq. (\ref{br36}) for the partial cycle amplitudes there appear new $2s$ poles (in the addition
to a slightly displaced the pole $\lambda = i$ of the bare Zwanzig model). Correspondingly, in such a case, the initial state evolution is described by superposition of $(2s +1)$  exponential terms. For the sake of simplicity let us consider the case $s=1$. Then the 3 poles are 
\begin{eqnarray}
\label{br40}
\lambda _1 = i\left (1 + \eta + \frac{3}{2}\eta ^2 + ....\right )\, ;
\lambda _{2, 3} = \pm i \left (\eta ^{-1/2} \mp (1/2) + ...\right ) 
\, .
\end{eqnarray}
At the critical value of the deformation parameter, $\eta _c$ the poles $\lambda _2$ and $\lambda _3$ become complex conjugated and are merged together forming a single second order pole. If $\eta \ll \eta _c$, then $|\lambda _1| \ll |\lambda _{2 , 3}|$, and the contribution of the new poles is small. As a result, we find in the agreement with Eq. (\ref{br28}), that the initial state amplitude decays almost exponentially with its decrement close to $\Gamma $. 
When the parameter, $\eta $, grows the decrement constant increases. The initial state population exhibits the damped oscillations when
the complex conjugated poles appear.  In this case, for the initial cycle $k=0$ its time evolution becomes non-exponential. The level population decay can be characterized by the following time-dependent effective decrement
\begin{eqnarray}
\label{br41}
d_s(t) = - \frac{d (log(a_s^{(0)}))}{d t}
\, .
\end{eqnarray}
The described above spectrum deformations create also the poles on the real axis, which result in oscillations of the partial amplitudes in  the negative local time region, $t < 2k\pi $. This a bit counter-intuitive  backward evolution (which, as it should be, disappears at $\eta =0$  is shown in the Fig.4 
\begin{figure}
  \begin{center}
    \includegraphics[height=3.5in, width = 225bp]{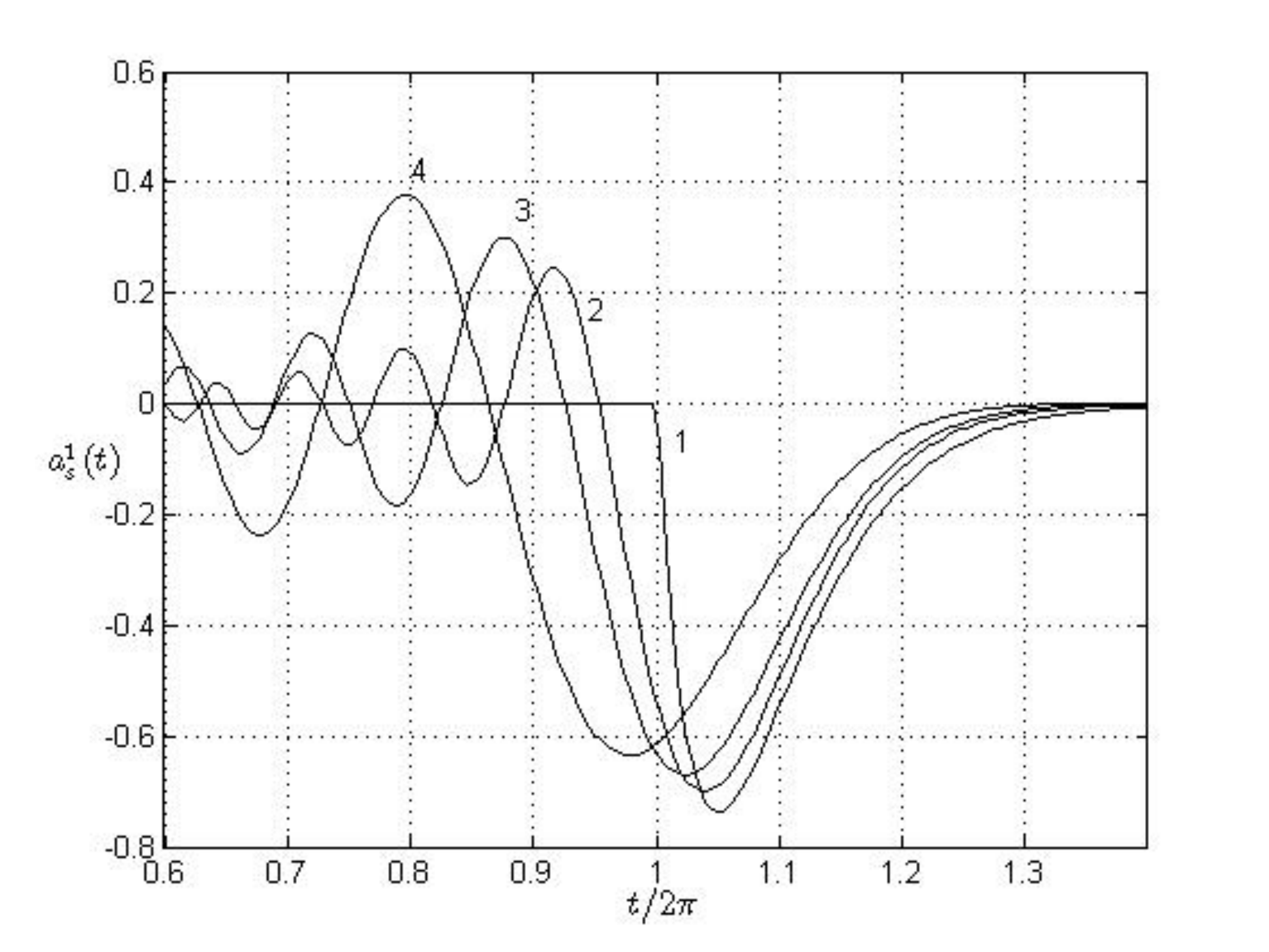}
    \caption{Backward evolution of the initial state amplitude for the cycle $k=1$. The reservoir posses the increasing inter-level spacings
according to the Exp.(\ref{br29}). $C^2=1$. The entering Eq. ({\ref{br29}) parameters: $b= 0$, $a = 0\,;\, 0.15\, ;\, 0.25\, ;\, 0.35$ (curves 1-4, respectively).}}
  \label{4}
  \end{center}
\end{figure}
From the all above examples we conclude that the crossover from regular to quasi-chaotic behavior is strongly dependent on the spectrum of the $S R$ mixed states. The number of the cycles with regular dynamics increases when the spectrum approaches to the equidistant one. 
In the case of the special spectrum deformation (\ref{br30}), the function $\phi (\lambda )$ can be represented as a trigonometric 
function of the angle $2\pi \varepsilon /(2s+1)$. The period of this trigonometric function is smaller than the basic period in the secular equation (\ref{br32}). This energy dependent angle in the deformation function $\phi (\lambda )$  leads to the satellites (weak components) in each of the partial cycle amplitude (apart from the basic Loshmidt echo characteristic for the bare Zwanzig model).
In time domain the satellites are shifted by $2\pi m/(2s+1)$ (where $ m = 1, ..... , s$ in each of the recurrence cycle. The intensities of the new components increase with increasing of  the deformation parameter $\delta $  in the Eq. (\ref{br30}). Alike the basic echo, 
the satellites are split into components, which start to overlap upon increasing of the cycle number. As a result, 
the critical cycle number decreases due to the mixing. As it is shown in the Fig. 5, the similar change in the evolution behavior results from the mixing deformations (\ref{br31}). For $\delta = 0$ (panel 5 a) the critical cycle number is given by the Eq. (\ref{br26}). For $\delta 
= 0.019$  and $\delta = 0.049$ (panels 5b and 5c) the satellite intensities are smaller than the basic components. 
The spectral triplets (which can be clearly visible in Fig 3) are accompanied by the new triplets in the fine structure of the absorption band. The decreasing of the critical cycle number, $k_c(\delta )$, corresponds to transfer of the satellites population for a given cycle into the next cycle. The mixing of the basic components and satellites produce the Loshmidt echo signal, consisting of many combined components with looking irregular intensity distributions. In a contrast with the bare Zwanzig model, the number of the echo signal components becomes several times greater than the cycle number. For $\delta = 0.079$ (panels 5d) the intensities of the basic components and satellites are already comparable, and the initial triplet structure is destroyed already in a few initial cycles, $k \leq 3$. Our numeric results presented in the Fig 5 demonstrate the change of the time evolution behavior under combined action of the dynamical (at $\delta =0$) and spectral mixing. If these processes are assumed to be mutually independent, the effective critical cycle number ${\tilde k}_c$ (for the reservoir consisting of three sub-lattices, $K=3$) is given by the following natural interpolation relation (confirmed by our numeric computations)
\begin{eqnarray}
\label{br42}
{{\tilde k}_c}^{-1} \simeq k_c^{-1} + 3 \delta 
\, .
\end{eqnarray}
The numerical calculations indicate that the Eq. (\ref{br42}) is satisfactory when $0 \leq \delta \leq 0.15$  and $1 \leq C^2 \leq 4$.
Note that the very possibility of such statistical description of the systems with spectral mixing is based on the ergodic hypothesis, which suggests that a single dynamical system visits all available states in its phase space, during the time evolution process (see, for example, 
\cite{KM10}, \cite{IZ90}, \cite{KB98}.
A comment concerning the interpolation formulas (\ref{br42}) is needed here. The fact is that for realistic parameters of nano-tubes or fullerens (or some other nano-particles) their dynamic evolution (essential for physical
applications) depends mainly on the $S R$  transitions into the limited number of states within the interaction region window. In this case, the spectral mixing affects the evolution dynamics only if the mixing occurs just in the interaction region (and all remaining system states  in the phase space are almost irrelevant). For the considered above types of the spectral deformations this condition is fulfilled if
$3\delta k_c >1$ (c.p., with our interpolation formulas (\ref{br42})). 
Although the cycle  period is energy dependent in the non-equidistant spectra, the regular evolution maintains in the sufficiently wide range of the deformation parameters. The recurrence cycles are destroyed only if the strong spectral mixing holds in the interaction region.

To conclude this subsection we would like to emphasize that the model spectral deformation  with $K=3$ sub-lattices 
we mainly considered in this subsection, is not merely a toy model to illustrate time evolution for the non-equidistant reservoir
spectra. For example, vibrational spectra of carbon nano-tubes \cite{KB98}, fullerene molecules \cite{SP02}, \cite{AC11},
or forming biological membranes lipids \cite{IY03} typically contain
indeed three vibrational energy levels progressions with nearly constant (but different in each of the progression) inter-level
spacings (in many lipids, see e.g., \cite{IY03}, these intervals are about $20\, cm$, and the corresponding period of the recurrence cycle is in pico-second time window). 
\begin{figure}
  \begin{center}
    \includegraphics[height=3.5in, width = 225bp]{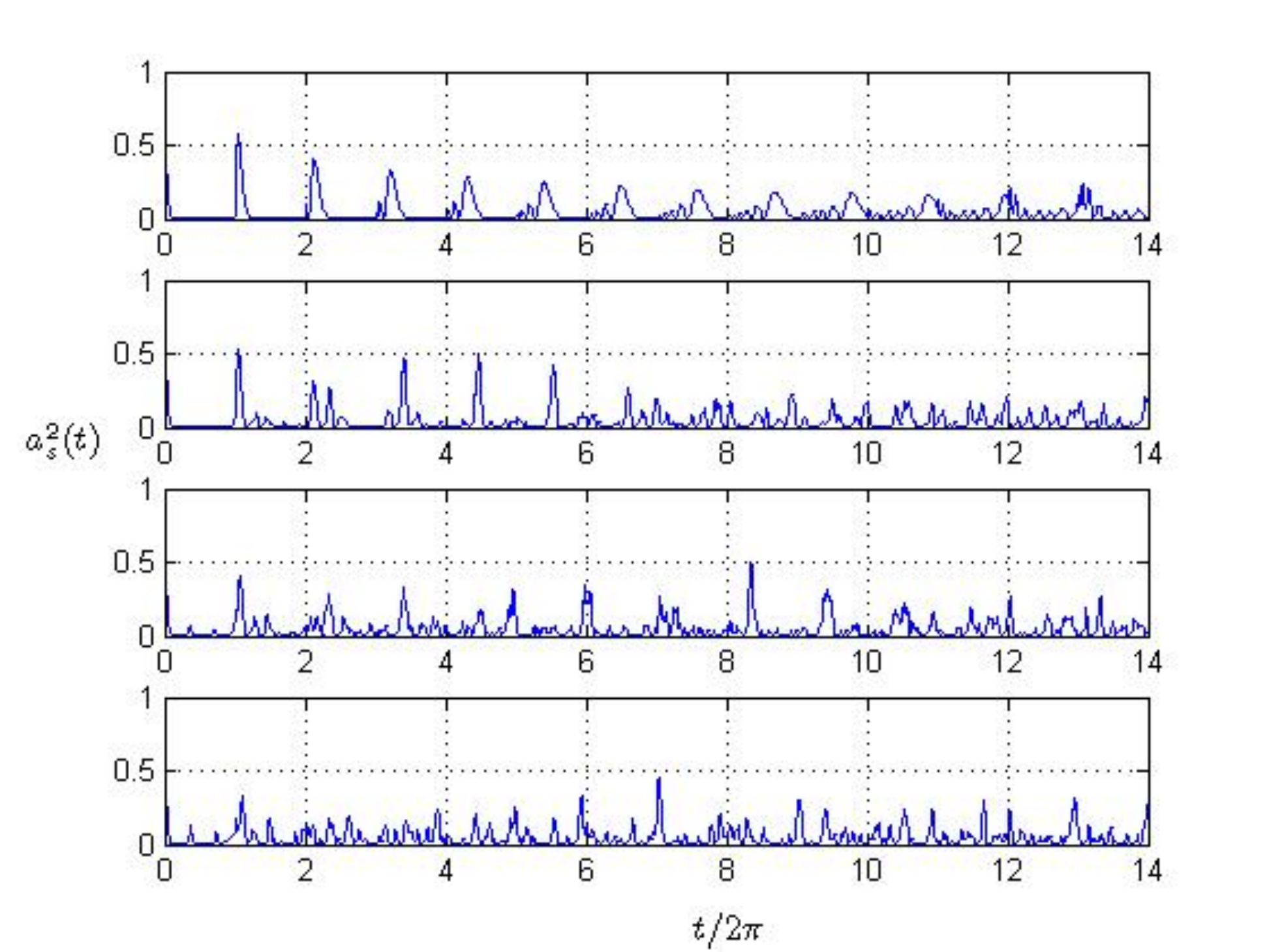}
    \caption{Time evolution of the initial state coupled to reservoir with spectral mixing. $C^2 =1$, $K=3$, $\delta = 0\, ;\, 0.019\, ;\, 
0.049\, ;\, 0.079$;  $a = 0\, ,\, 0.15\, ,\, 0.25\, ,\, 0.5$ (from the top to the bottom).} 
    \label{5}
  \end{center}
\end{figure}

\subsection{Ensemble of nano-particles at finite temperatures.}
\label{ensemble}

Usually in almost any realistic system there is not a single nano-particle but a composite consisting of many weakly coupled not-identical
nano-particles. Therefore applying the Zwanzig model approach in such a case, one has to deal with an ensemble of the reservoirs with somehow
distributed energy levels and their widths. The natural physical mechanism leading to the ensemble of the reservoirs is related to the collective vibrational excitations (phonons) coupled to the individual nano-particle displacements in the composite system. For instance the line width introduced ad hoc by hands in the section \ref{zwanzig} results just from the relative  displacements of the individual nano-particles. 
These displacements can be described in terms of the interaction between intra-molecular vibrations and phonons. Owing to this interaction the levels of the Hamiltonian (\ref{br2}) are displaced and acquired the finite width depending on the phonon quantum numbers. 
The phonon spectrum is limited by the Debye frequency characterizing the vibrations of the nearest neighbor particles. The Debye frequency 
$\Omega _D$ is typically by several times smaller than the intra-molecular frequencies in nano-systems under consideration (with their
characteristic scale frequency $\Omega _0$ scale frequency, and comparable with the thermal energy $k_B T$ at room temperatures
\begin{eqnarray}
\label{br43}
\Omega _D \simeq k_BT \ll \Omega _0
\, .
\end{eqnarray} 
Phonons yield to transitions between mixed internal vibration - phonon states
\begin{eqnarray}
\label{br44}
\langle n^\prime q^\prime | \leftarrow |n q\rangle  
\, ,
\end{eqnarray}
where $n$ stands for the internal vibration energy levels, and $q$ is phonon wave vector. Introducing phonon
creation and annihilation operators $b_q^+\, ,b_q$, 
the interaction of the arbitrary selected state $(n\, q)$ with the vibration-phonon reservoir is described by the Hamiltonian
\begin{eqnarray}
\label{br45}
H_{n q} = \varepsilon _{n q}^0 b_n^+ b_n b_q^+ b_q + \sum _{n^\prime q^\prime } \varepsilon _{n^\prime q^\prime }b^+_{n^\prime }b_{n^\prime }b_{q^\prime }^+ b_{q^\prime } +
\sum_{n^\prime q^\prime }V_{n q\, ;\, n^\prime q^\prime }(b_n^+b_{n^\prime } + b_n b_{n^\prime }^+)(b_q^+b_{q^\prime } + b_q b_q^+)  
\, .
\end{eqnarray}
To retain the well-defined absorption band, the phonon induced shifts and broadening should be smaller than the characteristic  inter-level spacings in the intra-molecular spectrum. The weak vibration-phonon coupling condition reads 
\begin{eqnarray}
\label{br46}
|V_{n q ; n^\prime q^\prime }| \ll |\varepsilon _n^0 - \varepsilon _{n^\prime }^0|
\, .
\end{eqnarray}
The weak vibration-phonon coupling is often realized in molecular crystals. In the opposite case of strong vibration-phonon coupling, 
the fine structure of the absorption band is smeared out. This limit can be treated theoretically  within the conventional theory of 
radiation-less transitions in systems with continuous spectra. When the condition of (\ref{br46}) is fulfilled, the integral over the continuous spectrum can be represented as the sum of the corresponding integrals over all individual intervals with one simple pole within each interval
 \begin{eqnarray}
\label{br47}
\sum_{n^\prime q^\prime } \to \sum _{n^\prime }\int d\varepsilon _q \rho (n^\prime , \varepsilon _q)
\, ,
\end{eqnarray}
where $\rho (\varepsilon )$ is phonon spectral density. Then we can derive the secular equation
\begin{eqnarray}
\label{br48}
F_{n q} = \varepsilon - \varepsilon _n^0 - \int \frac{V_{n q}^2(u)}{u - \varepsilon _{n q}^0}\rho (u) du =0
\, ,
\end{eqnarray}
where we denote 
\begin{eqnarray}
\label{br49}
V_{n q}(u) = V_{n q; n^\prime q^\prime }\delta (u - \varepsilon _{n^\prime  q^\prime }^0 )
\, .
\end{eqnarray}
Following Van Hove general formalism for the perturbation theory in systems with continuous spectra \cite{HO55},
and using exact relation
\begin{eqnarray}
\label{br50}
lim _{\eta \to 0} \int \Phi (x)(x + i \eta )^{-1} d x = - i \pi \Phi (0) + P\int \Phi (x) x^{-1} dx 
\, 
\end{eqnarray}
($P$ means the principal part of the integral over the single interval), we find that the pole $\varepsilon _{n q}^0$ in the lower 
complex half-plane determines the level width, while the sum  of the principal parts leads to the displacements of the intra-molecular 
energy levels
\begin{eqnarray}
\label{br51}
\varepsilon _{n q} =  \varepsilon _n^0 + D_{n q} - i \gamma _{n q}\,
;\, D_{n q} = \sum _{n^\prime }P\int \frac{V_{n q}^2(u)}{u - \varepsilon _{n^\prime q}^0}\rho (u) du\, ;\,
\gamma _{nq} = \pi V_{nq nq}
\, .
\end{eqnarray}
Eq. (\ref{br51}) defines complex eigen-values of the eigen-states $|n  q\rangle $. 
Due to vibration-phonon interaction a single intra-molecular level splits into the band with half-width $\gamma _{n q}$. According to the weak
coupling condition (\ref{br46}) the bands are not overlapped, and the vibration-phonon states are approximately orthogonal. 
With found from (\ref{br51}) values for the level displacements and broadening, the effective intra-molecular Hamiltonian matrix (with small imaginary parts of the diagonal elements resulted from vibration phonon coupling) can be written for a given value of the phonon wave number
$q$ as
\begin{eqnarray}
\label{br52}
H = \left (
\begin{array}{ccccccc}
\varepsilon _{2q} - i\gamma _{2q} & 0 & 0 &C_{2s} & 0 & 0 & .... \\
0 & \varepsilon _{1 q} - i\gamma _{1 q} & 0 & C_{1s} & 0 & 0 &...\\
0 & 0 & \varepsilon _{0 q} - i \gamma _{0 q}  &C_{0s} & 0 & 0 & ....\\
C_{2s}  & C_{1s} & C_{0s} &\varepsilon _{sq} - i\gamma _{sq} & C_{-1s} & C_{-2s} & .... \\
0 & 0 & 0 & C_{-1s} & \varepsilon _{-1q} - i \gamma _{-1 q} & 0 & 0 \\
0 & 0 & 0 & C_{-1s} & 0 & \varepsilon _{-2 q} - i\gamma _{-2q} & 0 \\
....& ....& ....& ....& ....& ....&...\\
\end{array}
\right ) 
\, , 
\end{eqnarray}
where 
\begin{eqnarray}
\label{br53}
\varepsilon _{s q} =  \Delta _0 + D_{s q}\, ;\, \varepsilon _{n q} = \varepsilon _n^0 -  D_{n q}
\, ,
\end{eqnarray}
and in the weak coupling limit
\begin{eqnarray}
\label{br54}
\gamma _{n q} \ll |\varepsilon_{n q} - \varepsilon _{(n+1) q}|
\, .
\end{eqnarray}
If characteristic optical excitation time is shorter than inverse Debye frequency and longer than period of intra-molecular vibrations, the initial state (prepared by the vertical optical pumping as it is schematically shown in the figure \ref{1}) is in the frozen phonon configuration and includes the set of phonon states determined by their equilibrium distribution in the ground state $\rho _g(\varepsilon _q, T)$. Then, according to the random phase principle, this distribution is random and can be characterized by the mean value and dispersion. Therefore, in order to describe the ensemble of nano-particles the Hamiltonian matrix (\ref{br52}) should be averaged over this distribution. In the weak coupling limit (\ref{br54}) we can introduce independently energy mean values and level widths of intra-molecular states  
\begin{eqnarray}
\label{br55}
{\bar {\varepsilon }_n(T)} = \int \varepsilon _{n q}(\varepsilon _q)\rho _g(\varepsilon _q ,T) d \varepsilon _q \, ;\,
{\bar{\gamma }_{n}}(T) = \int \gamma _{n q}(\varepsilon _q)\rho _g(\varepsilon _q , T) d\varepsilon _q
\, ,
\end{eqnarray}
and the corresponding dispersions
\begin{eqnarray}
\label{br56}
{\bar {\delta }}_n^2(T) = \langle (\varepsilon _{n q} - {\bar {\varepsilon }}_n)^2\rangle _\rho \, ;\,
\gamma _{n q} \ll |\varepsilon_{n q} - \varepsilon _{(n+1) q}|
\, .
\end{eqnarray}
From the other hand the fluctuation-dissipation theorem (see, for example the classical textbook \cite{LL77}, or its more
advanced version in \cite{YS94}) states that the temperature dependent  dispersion is given by
\begin{eqnarray}
\label{br57}
{\bar {\delta }_n}^2(T) = {\bar {\delta }_n^2(0)} \coth \frac{\varepsilon _q}{k_B T}
\, .
\end{eqnarray}
where ${\bar {\delta }_n(0)}$ is the width of the ground state distribution at $T=0$. Eqs. (\ref{br55}) - (\ref{br57}) suggests the following steps to describe an ensemble of nano-particles at finite temperatures: (i) to replace real-valued intra-molecular eigen-frequencies of the individual nano-particle  by the complex-valued counterparts for the ensemble of nano-particles, taking into account the weak vibration-phonon coupling;  (ii) to average over the set of  the ground  state configurations ( i.e. over random distribution 
with temperature dependent widths). The latter procedure to replace the mean values (\ref{br55}) by 
\begin{eqnarray}
\label{br58}
{\tilde {\varepsilon }_n} = {\bar {\varepsilon }_n} - i {\bar {\gamma }_{n}} + f_n
\, ,
\end{eqnarray}
where $f_n$ is a random variable with the zero mean value and the dispersion ${\bar {\delta }_n}^2$. This Eq. (\ref{br58}) assumes that
there are two mechanisms for the line broadening. Homogeneous broadening results from the transitions (\ref{br44}), whereas the non-homogeneous broadening arises from the temperature dependent distribution of phonon configurations in the ground state. As a result, the fine structure of the absorption band is not changed upon non-homogeneous broadening if
\begin{eqnarray}
\label{br59}
{\bar {\delta }_n} \ll |{\bar {\varepsilon }_{n}} -  {\bar {\varepsilon }_{n+1}}|
\, .
\end{eqnarray}
It is instructive to compare the absorption band structures for a single particle and their ensemble.
For a single particle the absorption band has the Lorentzian shape, while due to non-homogenous broadening (occurring as a result of the ensemble averaging) it acquires the Gaussian shape.  If the values ${\bar {\gamma }_n}$ and ${\bar {\delta }_n}$  are comparable the fine structure components have a rather complicated shape consisting of the Gaussian central part and Lorentzian tails. 
Let us stress that in the ensemble of nano-particles decay of the reservoir states (due to vibron-phonon interactions) suppresses the Loschmidt echo signal. Indeed as we said already above, to study the initially prepared state evolution in the ensemble of nano-particles, one has to average the initial state amplitude over the random distribution of the phonon configurations. This procedure leads to the broadening of the Loschmidt echo components. Then the number of observable recurrence cycles decreases. Only a finite number of the reservoir states is essential for the time evolution of the initial state, and this number is on the order of the $S R$ interaction region width, $\Gamma $. In fact for the Zwanzig model with finite number of the reservoir states the time evolution is almost the same as that for the infinite reservoir systems 
\cite{BG09}.

Note to conclude this subsection that similar results can be obtained within the random matrix formalism widely 
used in nuclear \cite{ME68} - \cite{PW07} and molecular \cite{DF00} spectroscopy.

\section{Reservoir states evolution}
\label{reservoir}
In the macroscopic systems the continuous reservoir states play a passive role,
merely as a sink for the population of the initially prepared state. The reason is evident, there are no reverse $R \to S$
transitions since the recurrence cycles periods are infinitely long. By contrast in nano-systems possessing reservoirs 
with discrete dense spectra, the reservoir enters actively into the system dynamics.
The states of the reservoir with a dense discrete spectrum are periodically occupied in the direct $S \to R$ transitions and then
depleted in the reverse $R \to S$ transitions. Synchronicity  of the reverse transitions leads to the double resonance phenomenon at the reservoir eigen-frequencies. These double resonances hold (and can be observed) even when the Loshmidt echo components are overlapped,
therefore the double resonances provide a convenient and powerful method for experimental studies of the ultra-fast energy redistribution processes \cite{HO07} - \cite{AS04}.

According to the Heisenberg equations of motion (\ref{br15})
\begin{eqnarray}
\label{br60}
a_n(t) = - iC_n \int _{0}^{t}\exp (i n(t-t^\prime ) a_s(t^\prime ) dt^\prime
\, 
\end{eqnarray}
or in terms of the partial cycle amplitudes
\begin{eqnarray}
\label{br61}
a_n(t) = \sum _{k=0}^{[t/2\pi ]}a_n^{(k)}
\, .
\end{eqnarray}
In the bare Zwanzig model, these amplitudes have the form \cite{BG09b}
\begin{eqnarray}
\label{br62}
a_n^{(k)} = i \frac{C}{2k \Gamma }\exp (-i \beta _n \tau _k) \int _{0}^{\tau _k} x L_{k-1}^1(x)\exp (-s_n x) dx
\, ,
\end{eqnarray}
where
\begin{eqnarray}
\label{br63}
\beta _n = \frac{n}{2\Gamma }\, ;\,
s_n = \frac{1}{2}\left (1 - i \frac{n}{\Gamma }\right )
\, .
\end{eqnarray}
Each of the amplitude in the expression (\ref{br62}) consists of the periodic (harmonic) and non-periodic components. At the end of each cycle, the non-periodic component is exponentially small while the periodic component, ${\tilde {a}}_n^{(k)}(t)$ has  the constant magnitude
\begin{eqnarray}
\label{br64}
{\tilde {a}}_n^{(k)} = C\frac{2\Gamma }{\Gamma ^2 + n^2} \exp (-i(nt - k(\pi + \varphi _n)))
\, ,
\end{eqnarray}
where the phase shift $\varphi _n $
\begin{eqnarray}
\label{br65}
\varphi _n = tan^{-1} \frac{n}{\Gamma }
\, .
\end{eqnarray}
The amplitude of the population in the $n$-th state of the reservoir is the sum of the partial amplitudes (\ref{br64}) for all previous
cycles
\begin{eqnarray}
\label{br66}
a_n(2(k+1)\pi ) \simeq \sum _{k^\prime =0}^{k} {\tilde {a}}_n^{(k^\prime )}(4(k - k^\prime + 1)k_c)
= i C (-1)^{k+1}\frac{(\Gamma + in)^k}{(\Gamma  - in)^{k+1}} \, , k < k_c
\, .
\end{eqnarray}
Then from (\ref{br66}) we conclude that the magnitudes of the reservoir state amplitudes at the end of the non-overlapped cycles are $k$-independent. Their form is given by the Lorentzian with the half-width  $\Gamma $
\begin{eqnarray}
\label{br67}
|a_n (2k\pi )|^2 =\frac{\Gamma }{\pi (\Gamma ^2 + n^2)} 
\, .
\end{eqnarray}
The phase shift $\varphi _n$ (\ref{br65}) by its meaning is the additional (with respect to standard dynamic phase ) geometric phase shift which results from each population exchange event. Since the number of exchange events is equal to the cycle number, the total shift is given by
\begin{eqnarray}
\label{br67a}
arg (a_n (2k\pi )) = k\pi + (2k+1) \varphi _n 
\, .
\end{eqnarray}
The periodic (harmonic) components describe the motion of a point in the phase space along a circle with the radius, $R_n = {\sqrt {\Gamma ^2 +
n^2}}$, which is independent of the cycle number. The long sequence of the rotations by the angles (\ref{br67}) incommensurable with $2\pi $, 
yields to a quasi-chaotic dynamic behavior. In the limit $k \to \infty $ all possible rotation angles densely cover the circle.
Based on our numeric results, the resonance time can be estimated as
\begin{eqnarray}
\label{br68}
\tau _{kn}^* \simeq \tau_{k0}^* \frac{\Gamma }{n} tan ^{-1}\frac{n}{\Gamma } 
\, ,
\end{eqnarray}
where $\tau _{0k}^* \simeq 4 k - 0.74 k^{1/3}$ corresponds to the resonance line for $n=0$ state in the $k$-th cycle.
The Eq. (\ref{br68}) is the condition for synchronic depopulation of the reservoir states with different numbers. 
The width of the distribution in the resonance times is proportional to the ratio, $(n/2\Gamma )^2$, which is small for the states contributing to the Loschmidt echo signal. It is worth to emphasize the difference between the shapes of the echo and double resonance lines. The width of the echo signal increases upon perturbations of the equidistant reservoir spectrum spectrum, whereas the double resonances 
occur irrespective of this property of the reservoir spectra. Thus the double resonances can be detected more easily than 
the Loschmidt echo signal which is smeared out due to large inhomogeneous broadening effects.

The amplitude of the resonance state, $a_0(t)$, always remains imaginary (similarly to the two-level system case).  The additional phase shift is zero, $\varphi _0 =0$, and the overturn (phase shift $\pi $) occurs along the imaginary axis between the points $\pm i C\Gamma ^{-1}$. 
In each cycle the partial amplitudes are changed by $\pm 2iC\Gamma ^{-1}$. The double resonance time, $\tau _{0k}^*$, is a characteristic time for a point in the phase space to pass through the center of the circle, when $a_0(\tau _{0k}^*) =0$. The global maximum of the double resonance signal is determined by the vanishing population for the maximal number of the reservoir states. 
The half-width of the double resonance line in the non-overlapping cycles increases slowly with increasing the cycle number. Our numeric
results suggest the following law 
\begin{eqnarray}
\label{br69}
\Delta \tau _{1/2} \simeq (32 k)^{1/3}\, ;\, 2 \leq k \leq k_c -1 
\, .
\end{eqnarray}
Owing to the overturns, the amplitude of the state $n=0$  has different signs in the odd and even cycles. In the odd cycles the population flow is directed from the reservoir toward the initial state, and the population of the resonance state decreases. By contrast, the population in the even cycles increases and becomes higher than the equilibrium  population given by the Lorentzian distribution (\ref{br66}). The population increases because the states with $n \simeq \Gamma $  more rapidly (in comparison with $n=0$ state) transfer the population into the initial state. The transitions from these $n \simeq \Gamma $ states produce the flow from the initial to the resonance state and lead to its super-equilibrium population. Therefore, the double resonance  transitions $n \to s \to n^\prime $ redistribute the population between the reservoir states. The double resonance transitions lead also to satellites in the double resonance spectra. At the end of the cycle $k=0$  the entire population of the initial state is transferred to the reservoir states ($a_s^{(0)}=0$) with the distribution (\ref{br66}). In the first cycle, the reservoir states populated in the zeroth cycle return the population back to the initial state with the initial rate proportional to $C^2$. Since the resonance time increases with increasing of the reservoir level number $n$, different reservoir states are depleted in the different times. However, the reservoir states are not depleted completely during the cycle, and therefore the amplitude of the initial state is not totally recovered.

Numerically we find that in the cycle $k=1$ the maximum of population amplitude $max |a_n|^2 \simeq 0.46$ is at $\tau _{s1}^* =2$. The states 
$n > \Gamma $ are depleted at $\tau _{n1}^* < \tau _{s1}^*$, when the population of the initial state is still increasing, and the states 
$n < \Gamma $ are depleted when  this population decreases. At the end of the cycle, the population is distributed over the reservoir states, as it is dictated by the Eq. (\ref{br66}). The similar behavior of the population redistribution occurs in the subsequent cycles. 
We illustrated this behavior for the cycle $k=2$ in Fig. 6.
\begin{figure}
  \begin{center}
    \includegraphics[height=3in, width = 225bp]{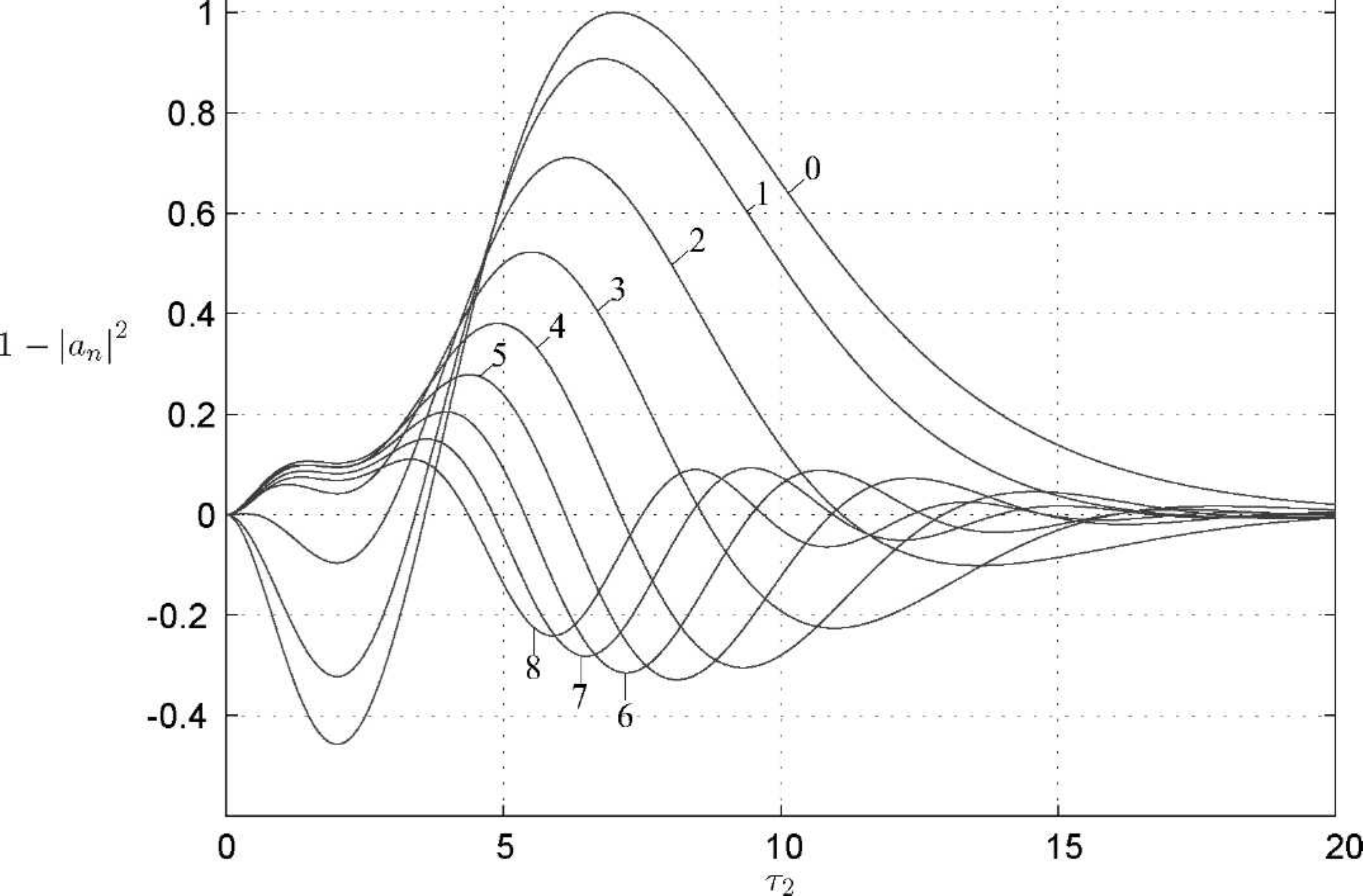}
    \caption{Redistribution of the initial state population over the mixed reservoir states, during the second recurrence cycle, $k=2$. The numbers of the reservoir states are shown on the curves.}
    \label{6}
  \end{center}
\end{figure}
One can find also that apart from the main double resonance line caused by the overturn, at $k \geq 2$ there arise $k-1$  
weak lines (satellites) which are associated with the oscillations of the initial state amplitude. The number, intensity and width of the satellites increase upon increase of the  cycle number. Nevertheless this change in the satellite spectrum does not affect the main double resonance, which intensity and width depend only weakly on the cycle number. As it is shown in Fig. 6, the satellites leave the main resonance behind in the states $n < \Gamma $, and they occur after the main resonance line in the states $n > \Gamma $. In the states $n \simeq \Gamma $  there are the both satellites, advanced and retarded ones.

The initial state plays the role of the intermediate energy level in the redistribution process between the reservoir states. The sign of the population flow changes $k$ times during the cycle depending both on the amplitudes of the population and phases. As a result, the dynamics of the redistribution is not described by the detail balance conditions. The fine structure of the Loschmidt echo signal is associated with the deviations of the reservoir state populations from the equilibrium distribution (\ref{br66}). In terms of the phase shift the condition 
for the Loschmidt echo mixing $k > k_c $ means that the phase shifts for all reservoir states $n \geq 1$ are larger than $2\pi $  when 
$k > k_c$. 
Mixing affects also the shape of the double resonance signal. It occurs the double resonance line width given by Eq (\ref{br69}) becomes comparable with the state under consideration vibration period, $2\pi /n$. The critical cycle number corresponding to this condition drops rapidly when the state number  increases 
\begin{eqnarray}
\label{br70}
k_n \simeq \frac{1}{4}\left (\frac{k_c}{n}\right )^3 
\, .
\end{eqnarray}
The criterion (\ref{br70}) defines also crossover from the regular to quasi-chaotic dynamics in the $n$-th reservoir state evolution. When
$k > k_n$ , the intensities of the main lines and satellites are comparable. Our numerical computations also support quasi-random dynamics in the cycles $k > k_n$. The computed trajectories in the phase space $a_n$ and $da_n/dt$ look as strongly entangled and fill the entire phase space available for a given reservoir state $n$.

\section{Other applications of the Zwanzig model}
\label{applications}

The purpose of this section is to illustrate how the bare Zwanzig model (or its generalization) approach can be applied
to describe quantum dynamics of various experimentally studied (and therefore interesting for theorists) nano-systems.

\subsection{Two-level systems coupled to reservoirs}
\label{two}
The main difficulty for experimental detection of the double resonances on the reservoir state frequencies is related to the fact that these frequencies are very close to the pumping frequency of the initial state. The required spectral resolution should be of the order of
the characteristic inter-level spacing, and it is not yet available. However the double resonances can be observed between two spectral well-separated intra-molecular states (see \cite{KM10}, or \cite{KN07} - \cite{RU09}). For such a case the initial (prepared by the optical pumping) and final (observed) states form a two-level system (TLS). Thus our aim in this subsection is to generalize described in the previous sections
Zwanzig model approach to study quantum dynamics of the TLS nano-systems. Note to the point that the 
TLS interacting with intra-molecular and phonon reservoirs is a convenient minimal model to investigate ultra-fast molecular processes
\cite{SS95} - \cite{MN08}. 

The dynamical problem for the TLS coupled to the reservoir with continuous spectrum has been investigated already in details 
(see, e.g.,  review paper \cite{LC87}, monograph \cite{BM94}, or our publication \cite{BK11}).
The specific feature of this model is the transition from coherent oscillations between the initial and final state to exponential decay of the initial state population when the transition matrix element between the TLS states becomes smaller than the matrix element for the TLS-reservoir coupling. In this subsection, we investigate the problem for the case of the intra-molecular reservoir with the 
equidistant spectrum (the bare Zwanzig model reservoirs). 
The time dependent wave function of the TLS coupled to the reservoirs is described by a column with 2 terms which are the wave functions of the initial and final configurations coupled to the reservoirs
\begin{eqnarray}
\label{bk71}
\Psi (t) = \left (
\begin{array}{c}
\Psi _L(t) \\
\Psi _R(t)  
\end{array}
\right ) 
\, . 
\end{eqnarray}
In own turns the functions $\Psi _L$ and $\Psi _R$ are superpositions of the initial (and final) TLS states ($\Phi _L$ and $\Phi _R$ respectively) coupled to the reservoir eigen-functions, $\Phi _{nL}$ and $\Phi _{nR}$, with time dependent amplitudes
\begin{eqnarray}
\label{br72}
\Psi _{L , R} (t) = a_{L , R}(t) \Phi _{L , R} + \sum _n a_{nL , nR}(t)\Phi _{nL , nR} 
\, .
\end{eqnarray}
The Hamiltonian, describing the TLS system coupled to the corresponding reservoirs, can be written in the following compact form
\begin{eqnarray}
\label{bk73}
H = \left (
\begin{array}{cc}
H_L & 0  \\
0 & H_R   
\end{array}
\right ) + \Delta {\hat {\sigma }_x} 
\, ,
\end{eqnarray} 
where $H_L$ and $H_R$ are the Zwanzig  Hamiltonian (\ref{br2}) for the $L$ and $R$ configurations, ${\hat {\sigma }_x}$ is the Pauli matrix,  
and $2\Delta $ is a level splitting in the isolated TLS (i.e., uncoupled to the reservoirs). The Hamiltonian matrix (\ref{bk73}) contains two non-zero columns and two non-zero lines of the coupling matrix elements for the both configurations and the diagonal matrix elements of the unperturbed TLS and reservoir states. Within the bare Zwanzig model, the both reservoir spectra (renormalized by the vibration-phonon interaction, described in Sec. \ref{generalizations}) (subsection \ref{ensemble}) are equidistant, and the coupling constants and decay rates are assumed to be  the same for all reservoir states
\begin{eqnarray}
\label{bk74}
E_{nL} + D_n = E_{nR} + D_n = n\, ;\, C_n = C\, ;\, \gamma _L = \gamma _R = \gamma _0\, ;\, \gamma _{nR} = \gamma _{nL} = \gamma
\, . 
\end{eqnarray}
The secular equation is
\begin{eqnarray}
\label{bk75}
F(\varepsilon ) = \left (\varepsilon + i \gamma _0 -\sum _{n}C_n^2(E_n + D_n - \varepsilon + i \gamma _n)^{-1}\right )^2 - \Delta ^2 =0
\, . 
\end{eqnarray}
This equation (\ref{bk75}) can be split into two equations for odd and even states. Performing explicitly the summation in the Eq.
(\ref{bk75}) we find these equations
\begin{eqnarray}
\label{bk76}
F^{\pm }(\varepsilon ) = \left (\varepsilon + i \gamma _0 \pm \Delta - \pi C^2\cot (\pi \varepsilon + i \pi \gamma \right ) =0
\, . 
\end{eqnarray}
Similarly we find the Heisenberg equations of motion, generalizing those (\ref{br15}) for the bare Zwanzig model  
\begin{eqnarray}
\label{br77}
i\frac{\partial a_L}{\partial t} = \Delta a_R + \sum _{n} C_n a_{nL}\, ;\,
i\frac{\partial a_R}{\partial t} = \Delta a_L + \sum _{n}C_n a_{nR}\, ;\, 
\end{eqnarray}
\begin{eqnarray}
\nonumber
i\frac{\partial a_{nL}}{\partial t} = E_n a_{nL} + C_n a_{L}\, ;\,
i\frac{\partial a_{nR}}{\partial t} = E_n a_{nR} + C_n a_{R}
\, .
\end{eqnarray} 
Solutions of these equations can be easily found, and can be represented in the following form
\begin{eqnarray}
\label{br78}
a_L(t) = \sum _{\varepsilon _n^+} \frac{\cos (\varepsilon _n^+ t)}{1 + \pi \Gamma + C^{-2}(\varepsilon _n^+ - \Delta )^2}\, ;\,
a_R(t) = i \sum _{\varepsilon _n^-}\frac{\sin (\varepsilon _n^- t)}{1 + \pi \Gamma + C^{-2}(\varepsilon _n^- - \Delta )^2}
\, .
\end{eqnarray} 
Transformation of (\ref{br78}) to the partial amplitude representation yields  
\begin{eqnarray}
\label{br79}
a_L(t) = \frac{1}{2}\sum _{k=-\infty }^{\infty } \left (a_k^{(+)} (t) + a_k^{(-)}(t)\right )\, ;\,
a_R(t) = \frac{i}{2} \sum _{k = -\infty }^{\infty }\left (a_k^{(+)}(t) - a_k^{(-)}(t) \right )\, ;\,
\\
a_k^{(\pm )}(t) = C^2\int _{-\infty }^{+\infty } d\varepsilon \exp (i \varepsilon (t - 2k\pi ))\frac{(\varepsilon \pm \Delta + i\Gamma )^{k-1}}
{\varepsilon  \pm \Delta - i \Gamma )^{k-1}}
\, .
\end{eqnarray} 
Calculating the contour integral we end up with the explicit expressions for the partial cycle amplitudes
\begin{eqnarray}
\label{br80}
a_L^{(0)} = \exp (-(\Gamma + \gamma _0)t) \cos (\Delta\, t)\, ;\,
a_L^{(k)}(\tau _k) = a_{s0}^{(k)}(\tau _k)\exp (-2k\pi \gamma )\cos \left (\frac{\Delta }{2\Gamma }\tau _k\right )
\exp \left (- \frac{\gamma _0 }{2\Gamma }\tau _k\right )
\, ,
\end{eqnarray} 
and
\begin{eqnarray}
\label{br81}
a_L^{(0)} = \exp (-(\Gamma + \gamma _0)t) \sin (\Delta\, t)\, ;\,
a_R^{(k)}(\tau _k) = i a_{s0}^{(k)}(\tau _k)\exp (-2k\pi \gamma )\sin \left (\frac{\Delta }{2\Gamma }\tau _k\right )
\exp \left (- \frac{\gamma _0 }{2\Gamma }\tau _k\right )
\, ,
\end{eqnarray} 
where in the both cases $k \geq 1$, and $a_{s0}^{(k)}$ and $\tau _k$ are defined by the Eqs. (\ref{br36}) - (\ref{br37}).
The above equations (\ref{br80}), (\ref{br81}) include three mutually independent factors related to the transitions between the TLS states, between these states and the reservoirs, phonons with and without phonons are taking into the consideration.  The factorization of these phenomena is the main advantage of the partial amplitude representation.

With these formal results, on the TSL coupled to the equidistant spectrum reservoirs, in hands
we are in the position to proceed further, and to analyze the details of the quantum dynamics, and corresponding physical consequences.
At beginning of each cycle $k \geq  1$ the TLS states are depleted ($a_L^{(k)}(0) = a_R^{(k)}(0) =0$). The $L R$ transitions occur in the local time $\tau _k$ and depend on the total population of the both TLS states. As it concerns to the reservoir states they are populated in 
the agreement with Eq. ({\ref{br67}) independently of the populations of the individual TLS states. Reverse transitions from the reservoir states restore the TLS state population. As a result of this fact, the transitions between the TLS states are switched on again.  
Obviously, this effect is absent if the TLS were coupled to a reservoir with continuous spectrum. The time dependencies $a_L(t)$ and $a_R(t)$
given by the Eqs. (\ref{br79}) - (\ref{br81}) show that the regular dynamics region exists independently of the value of
the TLS splitting, $\Delta $. When $\Delta \ll \Gamma $, the main part of the initial population of the unperturbed left state $L_0$  is transferred to the reservoir states $\L_n$, within the Lorentzian contour with the half-width $\Gamma $. The rest of the population participates in the $L - R$ exchange. The effective (time dependent) rate constant for the $L_0$ -state decay in the cycle $k=0$  is
\begin{eqnarray}
\label{br82}
k_L = - \frac{d}{dt}\ln (a_L^{(0)}) = \Gamma + \Delta \tan (\Delta\, t)
\, .
\end{eqnarray} 
Consequently the amplitude of the final TLS state, $R_0$, has a maximum at $t_m = \Delta ^{-1} \tan (\Delta /\Gamma )$, so that the  rate constant for the $L - R$ transitions in the vicinity of $t_m$ is given by
\begin{eqnarray}
\label{br83}
k_{L - R} \simeq \frac{\Delta ^2}{\Gamma }
\, .
\end{eqnarray} 
In fact this Eq. (\ref{br83}) defines the rate for the dissipative tunneling. Similar characteristics are widely used in the theory of the systems strongly coupled with the thermal bath in the continuous spectrum case 
(see e.g., \cite{LC87}, and \cite{CL83} - \cite{YU96}). Note that the rate constant (\ref{br83}) is independent of the density of the reservoir states.  

The ratio of the amplitudes of the unperturbed $R_0$ and $L_0$ states the time $t_m$ is 
\begin{eqnarray}
\label{br84}
\frac{a_{L}^{(0)}(t_m)}{a_R^{(0)}(t_m)} \simeq \frac{\Delta}{\Gamma }
\, .
\end{eqnarray} 
Owing to the $L - R$ transitions and transitions from the TLS to the reservoir states the populations of $L_0$ and $R_0$ becomes
equal in the limit of $k \gg 1$. The total population of the both the states is independent of the TLS splitting, $\Delta $, and is determined by the rate of the intrinsic decay of the reservoir states. The average per cycle TLS population is
\begin{eqnarray}
\label{br85}
\langle |a_L^{(k)}|^2 + |a_L^{(k)}|^2\rangle _k = \Gamma ^{-1}\exp (-4k\pi \gamma )
\, .
\end{eqnarray} 
Eq. (\ref{br85}) indicates that the reservoir states are populated during the main part of the recurrence cycle period, so that their decay with the rate constant $\gamma $ determines the population of the TLS. When $\gamma = \gamma _0 =0$, and $\Delta /\Gamma \ll 1$, the average per cycle populations of the $L_0$ and $R_0$ states are described by the approximate relation 
\begin{eqnarray}
\label{br86}
\langle |a_{L , R}|^2\rangle _k \simeq (2\Gamma )^{-1} \left (1 \pm (1+ \alpha ^2)^{-1}J_0(4k\alpha (1+\alpha ^2)^{-1/3})\right )
\, ,
\end{eqnarray} 
where $\alpha = \Delta /\Gamma $, and $J_0$ is the zero order Bessel function. Eq. (\ref{br86}) shows that the average per cycle populations exhibits the damped oscillations as a function of the cycle number $k$. For $k\alpha \gg 1$  the populations of the $L_0$ and $R_0$  states become the same $\simeq 1/2\Gamma $.

Our numeric results, presented in the Fig. 7, are in the agreement with described above features of the TLS time evolution.
The above described oscillations suggest that the TLS states are dressed (i.e., mixed with the reservoir states). 
The transition probability between these dressed states $L_0 + \{L_n\}$ and $R_0 + \{R_n\}$ is about $\Gamma $  times smaller than that for the unperturbed TLS. The damping coefficient of these slow oscillations results from the exchange of the dressed states with the far placed  reservoir states $n \gg \Gamma $. For $\Delta > 2 \Gamma $   the oscillations disappear and the equal populations of the TLS states is established already in the cycle $k=0$.
\begin{figure}
  \begin{center}
    \includegraphics[height=3.5in, width = 225bp]{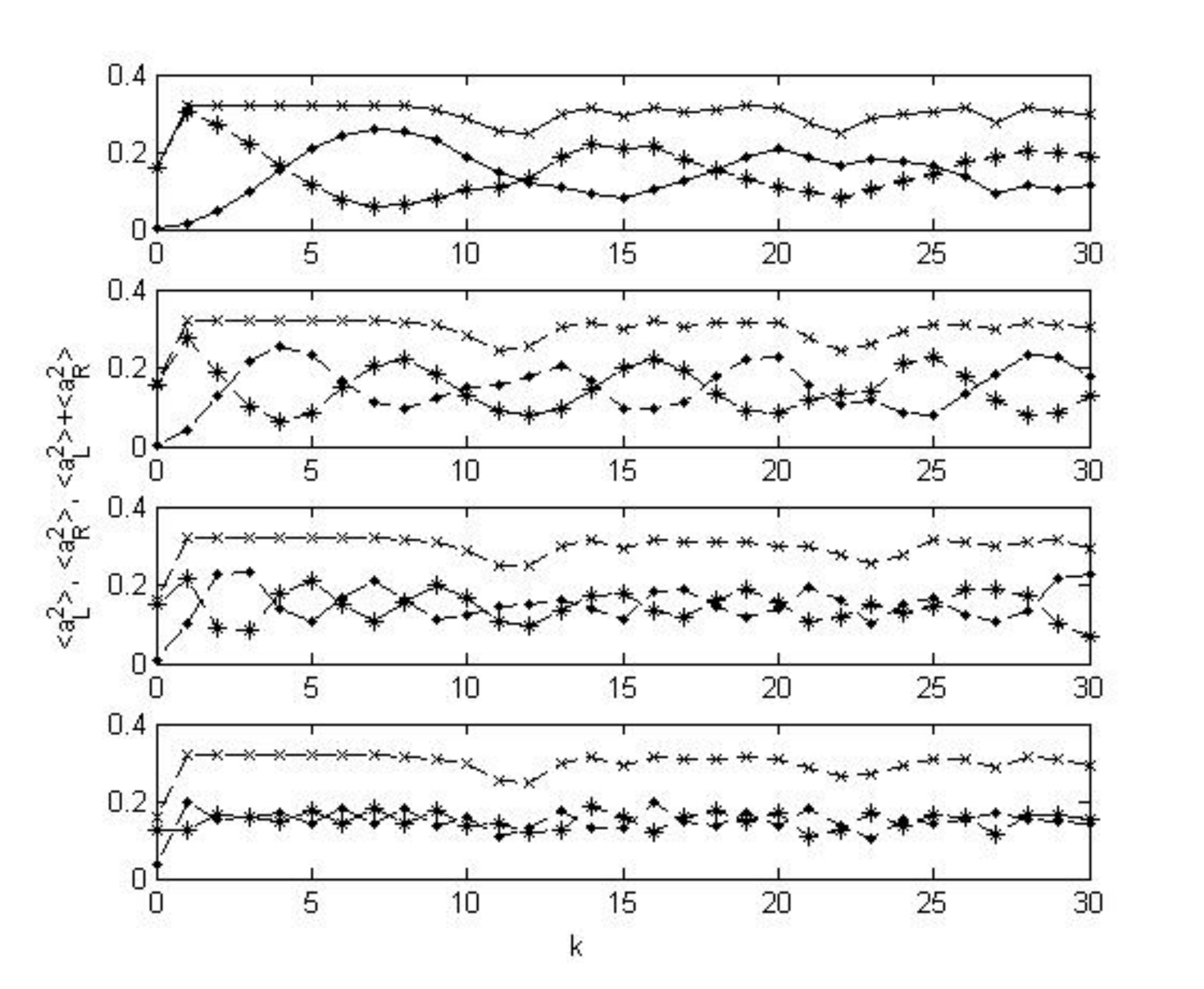}
    \caption{Variation of the average per cycle populations of $L_0$ and $R_0$, states and  the entire population of the TLS in the recurrence cycles. $C^2=1$, $\Delta = 0.4\, ,\, 1.2\, ,\, 2.7\, ,\, 7.4$ from the top to the bottom.}
   \label{7}
  \end{center}
\end{figure}

\subsection{Propagating vibrational excitations in nano-size chains}
\label{propagating}
All previous parts of our work were about systems which are nano-sized in three orthogonal directions. However there are several reasons behind a general interest to the systems which have nano-scales in only two or one directions. Most effects usually manifest themselves when at least one of the three system dimensions falls in the nano-sized regime. In such a regime a number of intrinsic system properties change, some qualitatively, 
others - quantitatively. We illustrate this conception investigating nano-sized chains, interesting for fundamental physics and applications.
Namely in this subsection we consider the finite length chain of the $(2N+ 1)$ coupled molecular or atomic  sites. All sites, except the central one at $n=0$, termed impurity site, are identical.
Each site can be only in the ground and in the first excited states. All highly excited states are neglected. This approximation is justified if all these states have sufficiently large excitation energy, and therefore are far from the low energy band formed by the coupled first excited states. The Hamiltonian of this model is 

\begin{eqnarray}
\label{br85}
H = E_0 Q_0^+ Q_0 + \Omega _0 \sum _{n} Q_n^+ Q_n + \sum _{\pm} C(Q_0^+ Q_{\pm 1} + Q_0 Q^+_{\pm 1})
+ J\sum _{n \neq 0, n \neq \pm 1}(Q_n^+Q_{n-1} + Q_n Q_{n-1}^+)
\, ,
\end{eqnarray} 
where $Q_0^+\, , Q_0$   are creation and annihilation operators of the impurity site $n=0$, and $Q_n^+\, ,\, Q_n$ are similar operators for all other sites, $n \neq 0$.  $E_0$ is the energy of the impurity excited state,  $\Omega _0$ is the energy in the center of the vibrational band, $C$ and $J$ are the matrix elements for the corresponding nearest neighbor couplings (all the same  for the uniform chain). 
The Hamiltonian (\ref{br85}) is written in the site (space) representation. It can be transformed into the quantum propagating wave  function representation, keeping separately one impurity state and the standing waves corresponding to two ($n < 0$ and $n > 0$) 
uniform half-chains. In this representation the band spectrum has the form
\begin{eqnarray}
\label{br86}
E_j = \Omega _0 + 2 J \cos \frac{\pi j}{N+1}\, ;\, j = 1 , 2 , .... N
\, .
\end{eqnarray} 
Eq. (\ref{br86}) represents the unperturbed spectrum of the half-chains which form the set of reservoir states. In contrast to the spectrum used in the bare Zwanzig model approximation, this spectrum becomes denser and denser upon approaching to the band boundaries, while
in the middle of the band, it is almost equidistant with the characteristic  inter-level spacing of the order 
$\Delta _0 \simeq 2\pi J/(N+1)$. Therefore, the recurrence cycle period $T$ is proportional to the chain length, $T \propto 2(N+1)/J$. 
In the propagating wave function representation, the coupling matrix elements $C_j$ are proportional to the amplitudes of the sites 
$n = \pm 1$ in the quantum $j$-th state
\begin{eqnarray}
\label{br87}
C_j = C (N+1)^{-1/2}\sin \frac{\pi j}{N+1}
\, .
\end{eqnarray} 
Eq. (\ref{br87}) indicates that the values of these coupling matrix elements are approximately constant near the center of the band
($j \simeq N/2$) and approaches to zero near the band boundaries. Alike the Zwanzig model discussed above, in the nano-scale chains, the dynamics of the vibrational excitation is governed by the parameter $\Gamma \propto C_{N/2}/\Delta _0$.
This parameter determines the number of reservoir states which essential in the impurity-chain interactions. For $0 \leq C \leq 1$ 
the impurity level lies within the band (\ref{br86}). 
In what follows for the sake of simplicity, we chose the zero energy level as $E_0$, and rescale all energies in $J$ units (i.e.,
$E=0\, , J=1$). Then the secular equation for the Hamiltonian (\ref{br85}) reads  
\begin{eqnarray}
\label{br88}
F(\varepsilon ) = (\varepsilon + E)D_N^2(\varepsilon ) - 2 C^2 D_N(\varepsilon ) D_{N-1}(\varepsilon ) = 0
\, ,
\end{eqnarray} 
where $D_N$ is  $C$ independent three-diagonal determinant, which is well known in the Jacobi matrix theory 
(see e.g., \cite{BE53}, \cite{OL74}), which is
\begin{eqnarray}
\label{br89}
D_N(\varepsilon ) = \frac{\sin (N+1) k}{\sin k}
\, ,
\end{eqnarray} 
where $\varepsilon = 2 \cos k$.
Eq (\ref{br88})) is reduced to two equations for the anti-symmetric and symmetric eigen-states of the chain with 
a single impurity site in the middle
\begin{eqnarray}
\label{br89}
D_N =0\, ,\, (\varepsilon + E)D_N(\varepsilon ) - 2 C^2 D_{N-1}(\varepsilon ) = 0
\, .
\end{eqnarray} 
For $C=0$  the levels of the uncoupled impurity-reservoir are doubly degenerated. For $E=0$   and odd $N$  values, the impurity level coincides with the reservoir level $\varepsilon ^0_{(N+1)/2}$. 
All essential physical ingredients of the excitation propagation along the nano-scale chain, can be illustrated in this particular case. 
Changing the variables in the secular equation (\ref{br88})
\begin{eqnarray}
\label{br90}
k = \frac{\pi }{2} - \alpha \lambda \, ;\, \alpha = \frac{\pi }{N+1}
\, 
\end{eqnarray} 
the secular equation (\ref{br88}) can be rewritten in the same form as the Eq. (\ref{br32})) for the generalized Zwanzig model
\begin{eqnarray}
\label{br91}
F(\lambda ) = f_2(\lambda )(\lambda - \Gamma f_1(\alpha \lambda )\cot (\pi \lambda )) = 0
\, ,
\end{eqnarray} 
where $f_1$ and $f_2$ are  the following corrective functions to the bare Zwanzig model 
\begin{eqnarray}
\label{br92}
f_1(x) = x \cot x\, ;\, f_2(x) = \frac{\sin x}{x}
\, .
\end{eqnarray} 
Thus, the dynamical problem for the chain with a single impurity site becomes equivalent  to the discussed above problem for the 
initially excited state (it is the impurity state in the chain problem) coupled to the reservoir with the non-equidistant spectrum and energy-dependent coupling matrix elements.  Assuming the following initial conditions for the recurrence cycle amplitudes
$a_0(0)=1$, and $a_k(0) =0$ we find the the Fourier series for the impurity state amplitude 
\begin{eqnarray}
\label{br93}
a_0(t) = 2\sum _{k=0}^{[N/2]} U_k(\lambda _k)\cos (2t \sin (\alpha \lambda _k))\, ;\,
U_k(\lambda ) = \frac{f_1^2(\lambda )}{(1 + \pi \Gamma )f_1^2(\lambda ) + (\alpha ^2 + \pi/\Gamma)\lambda ^2}
\, ,
\end{eqnarray} 
where
\begin{eqnarray}
\label{br94}
\Gamma = \frac{C^2(N+1)}{\pi (1 -C^2)}
\, .
\end{eqnarray} 
Replacing also time $t$ by the recurrence cycle time $\tau _k $
\begin{eqnarray}
\label{br95}
\tau _k = \frac{\sin (\alpha \lambda }{\alpha \lambda } t - 2k \pi
\, 
\end{eqnarray} 
we end up with the exact solution (\ref{br93}) describing impurity site excitation propagation along the nano-size chain.
It reduces to Eq. (\ref{br19}) when the function $f_1(\alpha \gamma ) \simeq 1$. For $C^2 < 1/2$ the leading contribution in the expansion
(\ref{br93}) results from the states of the almost equidistant spectrum (i.e., near the band center, where $k \leq \Gamma \ll N/2$). 
Since the states located near the band boundaries do not participate in the impurity-reservoir exchange, the evolution in this case resembles closely that for the bare Zwanzig model. According  with the Eq. (\ref{br94}), the coupling constant $\Gamma $ grows with increasing $C^2$.
Owing to the increase in $\Gamma $, the region of the regular dynamics increases and achieves the maximum at $C^2 =1$, 
where  the spectrum is nearly equidistant. 

For $0 \leq C^2 \leq 1/2$, the sum (\ref{br93}) can be transformed using the Poisson summation formula
\begin{eqnarray}
\label{br96}
a_0^{(k)}(\tau _k) = \frac{1}{\pi (1-C^2)} \int _{-\infty }^{\infty }\frac{\exp (i \lambda \tau _k)}{\lambda ^2 + \Gamma ^2 f_1^2(\lambda )}
\left (\frac{\lambda + i \Gamma f_1(\lambda )}{\lambda - i \Gamma f_1(\lambda )}\right )^k f_1^2(\lambda ) d \lambda
\, .
\end{eqnarray} 
The contribution of the compressed dense states located near the band boundaries increases when $C^2 > 1/2$. 

In Fig. 8 we present our numeric results to illustrate how the dynamics of the vibrational excitation propagation along the 
nano-chain depends on the coupling constant $C$. 
For $C^2 < 1/4$ dynamic behavior is similar to that for the bare Zwanzig model (Sec. \ref{initial}). The number of the components and the total width of the Loschmidt echo signal increases proportionally to the cycle number (panel 8a). The critical cycle number for the regular-chaotic crossover is
\begin{eqnarray}
\label{br101}
k_c \simeq (N+1)\frac{C^2}{1-C^2}
a_0(t) \simeq \sum _{l}(-1)^l ((1-C^2)(J_{2l(N+1)}(2t) + J_{2l(N+1) +2}(2t) + J_{2l(N+1)}(2t))
\, .
\end{eqnarray} 
However for $C^2 \geq 1/2$  dynamics is qualitatively different from that predicted by the bare Zwanzig model. Already for the small cycle numbers $k \geq 1$ (panel 8b) the impurity state population oscillates (instead of the exponential decay in the bare Zwanzig model). The oscillation amplitudes increase with $k$, whereas the intensity of the main component slightly decreases. Dynamics remains regular up to the cycle where the oscillating components become comparable with the main peak (panels 8c and 8d). Because the zeros of the Bessel functions of different orders are rationally independent, mixing of the partial amplitudes yields the irregular damped oscillations in $a_0(t)$. For $C^2 \simeq 0.4 $,  the critical number achieves the maximal value, and for larger values of $C^2$, $k_c$ decreases.  Therefore, the threshold for the quasi-stochastic irregular behavior increases in the interval $0 \leq C^2 \leq 1/4$, and it decreases for $1/2 \leq C^2 <1$. These two regimes correspond to the distinct mechanisms of randomization due to the mixing of the echo components. Namely: (i) cycle overlapping, considered
in section \ref{initial} and (ii) considered in this subsection enhancement of the irregular oscillation amplitudes at $k >k_c$. 
A more abrupt transition from regular (coherent) dynamics to stochastic one occurs after 
the first passage of excitation through an impurity site.
\begin{figure}
  \begin{center}
    \includegraphics[height=3.5in, width = 225bp]{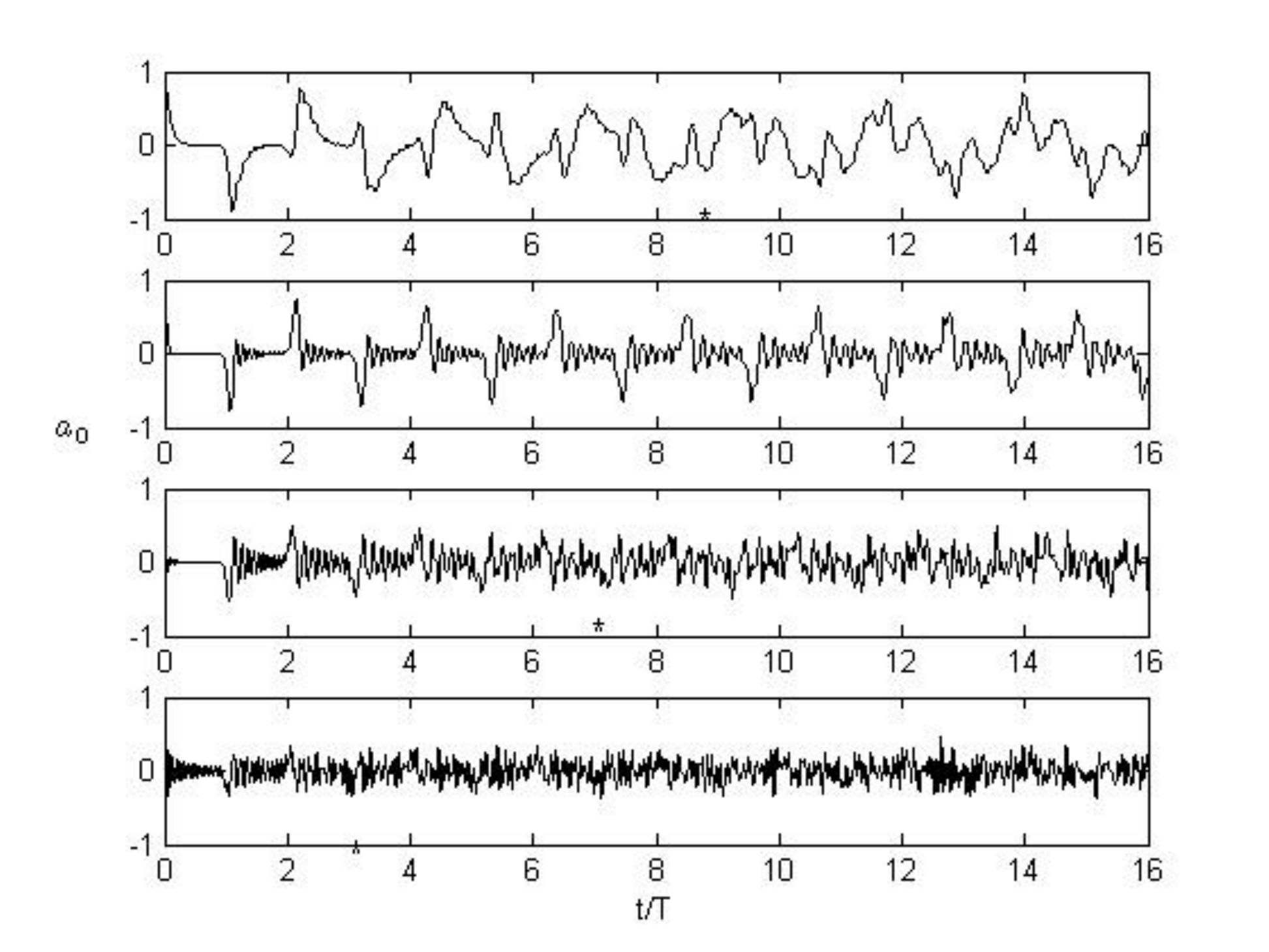}
    \caption{Dynamics of the population at the excited impurity cite located in the middle of  the finite $N=49$ size chain. $C^2 = 0.1\, ,\,
0.25\, ,\, 0.5\, ,\, 0.95$ from the top to the bottom. The critical cycle number is marked by the a star (for $C^2 = 0. 5$ and $k_c = 23$).}
    \label{8}
  \end{center}
\end{figure}

Our results for the symmetric nano-chain with a single impurity site in the center at $n=0$, can be easily generalized for a non-symmetric case, when the impurity site is located at $n \neq 0$. For such a case the time dependent amplitudes of the chain sites can be also expanded over the Bessel functions
\begin{eqnarray}
\label{br102}
b_n(t) = \frac{1}{1 -C^2}\sum _{m=0}^{\infty } J_{2m}(2t) S_{nm}(\Lambda )
\, ,
\end{eqnarray} 
where
\begin{eqnarray}
\label{br103}
S_{nm}(\Lambda ) = i\frac{C\Gamma }{\pi } \sum _{k=0}^{[N/2]}U_k(\lambda _k)\cos (2m \alpha \lambda _k)
\frac{\sin ((N+1 -n)\alpha \lambda _k)}{\sin ((N+1)\alpha \lambda _k)}
\, .
\end{eqnarray} 
The coefficients $S_{nm}$ are non-zero only in the narrow (on the order of $\Gamma $ and periodically repeated time intervals. Therefore,  
$b_n(t)$ similar to $a_0(t)$  is expanded over the partial amplitudes of recurrence cycles 
\begin{eqnarray}
\label{br104}
b_n(t) = \sum _{k} b_n^{(k)}(t - 2(N+1)k)
\, ,
\end{eqnarray} 
where the $k$-th partial amplitude is expressed in terms of the Bessel functions with their orders in the interval
$[k(N+1) + n \, , (k+1)(N+1) - n]$. Since the partial amplitudes depend on the combination $t -n$ , this interval determines also
the arrival $t_{nk}^{(1)}$ and departure $t_{nk}^{(2)}$   times for the located at the $n$-th site, excited impurity
\begin{eqnarray}
\label{br104}
k (N+1) + n = t_{nk}^{(1)} \leq t \leq t_{nk}^{(2)} = (k+1)(N+1) - n
\, ,
\end{eqnarray} 
where the interval, $t_{nk}^{(2)} - t_{nk}^{(1)}$ decreases upon increasing the impurity site $n$ number. As a result, the excitation propagates along the chain with almost constant speed (see Fig. 9). For the small coupling constant in the cycle $k=0$ there are two separate echo signals at times $t_{nk}^{(1)}$  and  $t_{nk}^{(2)}$ when the excitation wave passes through the $n$-th site.
\begin{figure}
  \begin{center}
    \includegraphics[height=3.5in, width = 225bp]{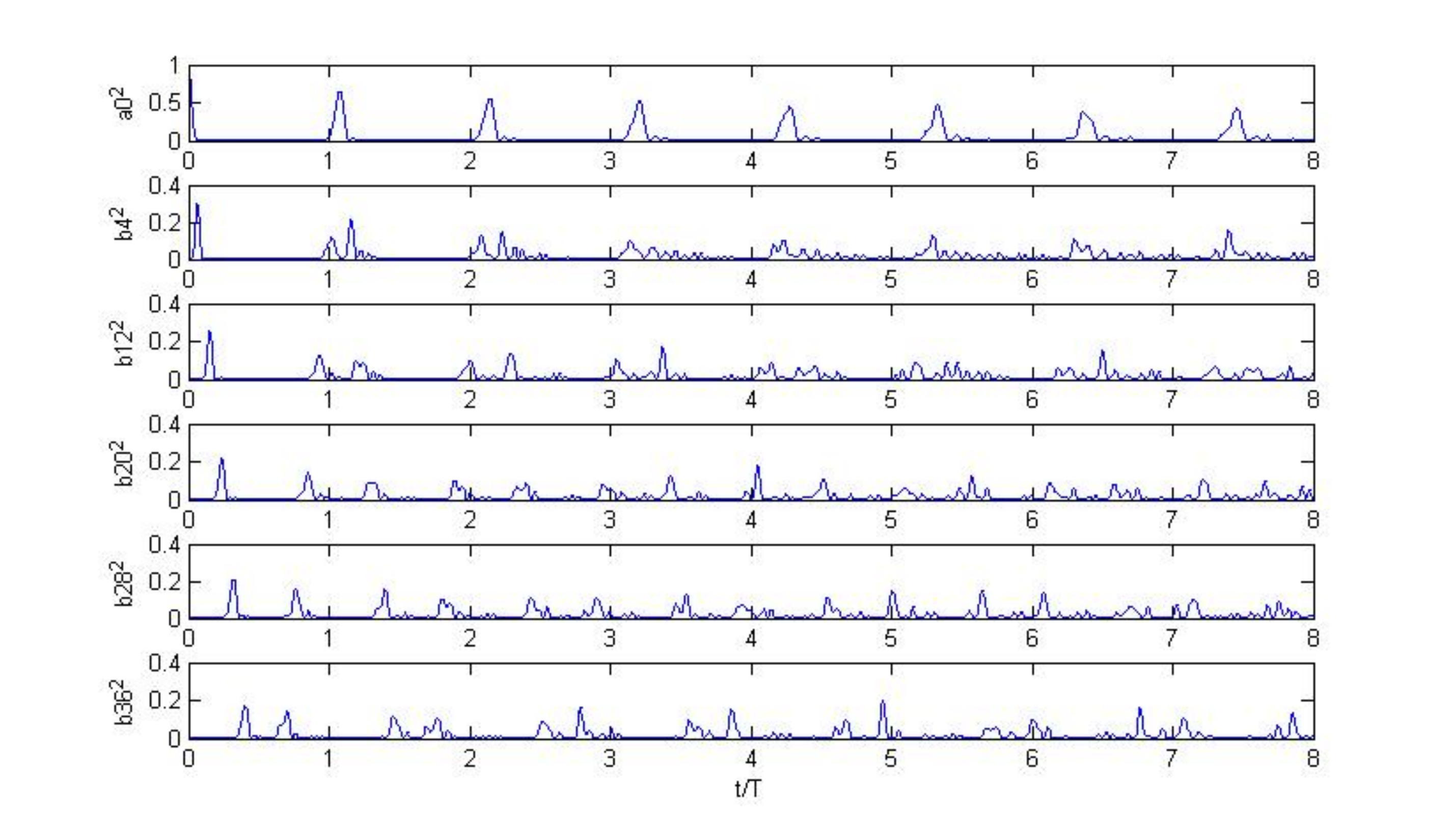}
    \caption{Impurity site excitation propagated along the chain. $C^2 = 0.5$, $N=49$. The cite numbers are indicated in the panels.}
    \label{9}
  \end{center}
\end{figure}
The excitation spreads over several sites and the spreading width is almost $n$-independent. For larger  $C^2$ values, the excitation propagates with the same (as for the weak coupling) speed but the intensity of the signal considerably decreases. Ultimately, for $C^2 \geq 1/2$  the width of the wave packet becomes of the order of the half-cycle time. In this limit the wave packet acquires the considerable width, the propagation is not ballistic anymore. The sites, $n \geq N/2$, are excited irregularly, whereas the impurity site dynamics remains regular. This phenomenon  means a sort of self-synchronization of the forward and backward (after reflection from the chain ends) waves when they are passing trough the impurity site. This process resembles the synchronicity of the reverse reservoir-initial state transitions we investigated in Sec. \ref{reservoir}

It is worth to note that the main findings in this subsections are in the agreement with experimental data \cite{WC07}, \cite{CW09},
manifesting that (i) in the uniform molecular nano-scale chain the vibrational excitation of the impurity site propagates almost ballistically for a long (on molecular scale) time and distances; (ii) the number of the recurrence cycles with almost regular behavior achieves its maximal value at the intermediate coupling constant $C^2 \simeq 0.4$.

\section{Conclusion}
\label{con}

Understanding all limitations of our simplified exactly solvable models, we nevertheless hope that the results collected in this review capture the essential peculiarities in nano-system dynamics. Namely, the dense discrete spectra characteristic for nano-particles and large-size molecules are responsible for appearance of recurrence cycles. By contrast to macroscopic systems with the continuous spectra, where the initial state population decreases monotonically  in time, the Loschmidt echo and double resonances arise in each recurrence cycle. The revivals in the time evolution makes it possible the emergence of new mechanisms of vibration stimulated processes in the single nano-particles and their ensembles as well. The regular behavior in the initial recurrence cycles transformed into chaotic-like long-time evolution is an inherent characteristic of nano-world. The combination of the deterministic spectra with chaotic-like long time dynamics has no analogs in the quantum dynamics of the macroscopic dissipative systems.  Owing to specific mechanisms of energy transfer and redistribution within the manifold of intra-molecular modes, the loss-free and distant energy transfer becomes possible.
The unique property of nano-systems is the possibility for periodical accumulation of   the energy on the selected vibrational modes. The new fields for experimental studies and applications of this unusual effect are opened.

\acknowledgements 
 
 We would like to thank the referee for carefully reading the manuscript and for his or her recommendations.

\end{document}